\documentclass[aps,twocolumn,prd,showpacs,preprintnumbers,nofootinbib]{revtex4}
\usepackage{amsmath}
\usepackage{graphicx}
\usepackage{dcolumn}
\usepackage{bm}
\usepackage{amssymb}
\usepackage{latexsym}
\usepackage{pifont}

\usepackage{pstricks}
\usepackage{color}
\usepackage{axodraw2}

\newcommand{\ie}{\emph{i.e. }}

\newcommand{\df}{\delta\phi}

\newcommand{\mean}[1]{\left\langle #1 \right\rangle}

\newcommand{\tin}{{t_\mathrm{in}}}

\newcommand{\dd}{\mathrm{d}}
\newcommand{\kk}{\mathbf{k}}

\newcommand{\xx}{\mathbf{x}}
\newcommand{\yy}{\mathbf{y}}

\renewcommand{\t}{\tau}
\newcommand{\e}{\mathrm{e}}

\newenvironment{pic}[3]{%
  \begin{minipage}[c][#2\unitlength][c]{#1\unitlength}%
  \begin{picture}(#1,#2) #3}{%
  \end{picture}\end{minipage}}

\def\Cross(#1,#2){
\put(#1,#2){
}
\put(#1,#2){
}
}

\begin{document}

\preprint{UTTG-16-06}

\title{A new diagrammatic representation for correlation functions in
the in-in formalism}

\author{Marcello Musso} 
\email{musso@physics.utexas.edu}
\affiliation{University of Texas at Austin, Department of
Physics - Theory group, 1 University Station C1608, Austin TX
78712-0269 USA}
\thanks{Present affiliation: \\
IRMP - CP3, Universit\'e catholique de Louvain, \\
Chemin du Cyclotron 2, B1348 Louvain-la-Neuve, Belgium}

\begin{abstract}
In this paper we provide an alternative method to compute correlation
functions in the \emph{in-in} formalism, with a modified set of
Feynman rules to compute loop corrections. The diagrammatic expansion
is based on an iterative solution of the equation of motion for the 
quantum operators with only retarded propagators, which makes 
each diagram intrinsically local (whereas in the standard case 
locality is the result of several cancellations) and endowed with
a straightforward physical interpretation. 
While the final result is strictly equivalent, as a bonus the 
formulation presented here also contains less graphs than other 
diagrammatic approaches to \emph{in-in} correlation functions.
Our method is particularly suitable for applications to cosmology.
\end{abstract}

\maketitle


\section{Introduction}

There has been an increasing interest in recent years towards the
study of non-linearities during the inflationary era.
It is very well known that inflationary theories predict the behavior 
of the inflaton and metric fluctuations to be fairly approximated by 
that of very light fields with nearly Gaussian correlation functions. 
Still, although severely constrained by observations \cite{wmap3}, the 
non-Gaussianity of the primordial cosmological perturbations is not 
ruled out. Some level of non-Gaussianity is indeed expected, as a 
consequence of the intrinsic non-linearity of the theory at least 
in the gravitational part of the action.
However, corrections to the linear behavior are suppressed by the
smallness of the couplings, which involve the Newton constant and
the slow-roll parameters \cite{Acquaviva, Maldacena, Bartolo}. 
This makes their predicted value so small that the detection of 
non-Gaussianity would be very difficult, at least for the simplest 
models of inflation when just one scalar field is involved.

Recently, it has been suggested \cite{Weinberg} that larger 
non-linear contributions to the statistics of the cosmological 
perturbations could in principle come from loop corrections to the 
correlation functions. The interest of studying loop contributions 
is that they contain powers of the scale factor -- which is growing 
exponentially -- integrated over the total duration of inflation.
When contributions beyond the tree level are taken into account, 
more powers of the coupling constant are included with the additional
interaction vertices; nevertheless, this suppression might be 
compensated by the integral over time of the scale factors in the 
loop. At the end of inflation, loop contributions could in principle 
grow large, modifying the tree-level behavior of the correlation 
functions. Even though in practice such integrals are not able to 
produce positive powers of the scale factor $a(\t)$ (which would 
easily overcome the powers of the coupling constants), but can grow 
at most as powers of $\log a(\t)$, some work is still needed
in that direction \cite{Sloth}.

Unfortunately, the mathematical aspect of computing quantum loop 
corrections in cosmology is more involved than in ordinary quantum
field theory. The reason is that the correlation function we want to
compute are true expectation values between the same \emph{in} vacuum
and not \emph{in-out} transition amplitude. Actually, the very same
concept of an \emph{out} vacuum is misleading in cosmology, since
there is not a moment when one can safely consider the interaction 
to be switched off.

When one wants to compute \emph{in-in} expectation values, Schwinger's 
formalism \cite{Schwinger,Mahanthappa,Keldysh,Chou,Jordan,Hu} has to be used. 
In this formalism, the time is thought to evolve along a close path,
from the remote past to some finite time (after the time at which 
expectation values have to be evaluated) and back. The evaluation of 
Feynman diagrams involves an extended set of rules, since two different 
fields $\phi_+$ and $\phi_-$ are introduced in the forward and
backward time branches respectively. As a consequence, one must
distinguish positive and negative vertices describing $\phi_+$ and 
$\phi_-$ interactions.
The propagator between two positive vertices is time ordered as usual,
while the one connecting negative vertices is anti-time ordered.
Finally, mixed propagators also exist, in which by definition $\phi_-$ 
is following $\phi_+$ in time.
This fairly complicated set of rules makes the computation of the 
expectation values rather cumbersome, due to the presence of many more 
terms than for the \emph{in-out} transition amplitudes in standard QFT.

A different formulation of the \emph{in-in} formalism was given in
\cite{Weinberg}. This approach is particularly suitable for 
applications to cosmology because it automatically takes care of
several cancellations in the late time behavior of the different 
contributions, which are not immediately apparent in the usual 
formulation. 
This is achieved writing each term of the perturbative series as a 
product of either commutators of two free fields or expectation 
values of two free fields. 
Unfortunately, lengthy calculations are still needed in order to 
obtain these reformulated contributions to the correlation functions.
Albeit straightforward, this procedure can be very long and 
involved, especially for high order terms.

The aim of the present paper is to use a set of modified Feynman 
rules to provide a graphical representation for this new formulation, 
in order to combine the useful features of this approach with the 
power and efficiency of the diagrammatic expansion. Of course, the
two formulations being two strictly equivalent ways of computing the 
same \emph{in-in} correlation functions, also the graphical representation 
outlined here will return the same final result, once all diagrams have 
been summed.

The plan of the paper is the following. In Section \ref{standardform} 
we review the basics of the \emph{in-in} formalism as described in
\cite{Weinberg}. In Section \ref{pertexp} we develop an alternative 
formulation leading to the 
same results. In Section \ref{graphrep} we provide a graphical 
representation of our procedure, and in Section \ref{Feyn} we make 
the connection with a modified version of Feynman's diagrams, whose
symmetry factor is discussed in Section \ref{Symmetry}.
Finally, we draw our conclusion in Section \ref{conclusion}.


\section{The \emph{in-in} formalism}
\label{standardform}

We consider the action for a self-interacting scalar field $\phi$ in
a fixed de Sitter space-time background. The metric for such space-time
is
\begin{equation}
  \dd s^2=\dd t^2-a^2(t)\dd\xx^2
\end{equation}
where the scale factor $a(t)$ evolves exponentially like $\e^{Ht}$,
so that the Hubble parameter $\dot a/a = H$ is constant.
This action reads
\begin{equation}
  S[\phi] = \int \dd x \,a^3 \!\left[
  \frac{1}{2}\partial_\mu\phi\partial^\mu\phi -
  \frac{1}{2}m^2\phi^2 - V(\phi)
  \right]
\end{equation}
giving, when variated with respect to $\phi$, the equation of motion
\begin{equation}
\label{eqmot}
  \ddot\phi + 3H\dot\phi - \frac{\nabla^2\phi}{a^2} 
  + m^2\phi + V'(\phi) = 0 \,.
\end{equation}

The retarded Green function for the Klein Gordon equation is the
solution of the inhomogeneous equation
\begin{equation}
  \left(\partial_t^2 +3H\partial_t -\frac{\nabla^2_x}{a^2} +m^2\right)
  \!G_R(x,x') = \frac{\delta(x-x')}{a^3}\,;
\end{equation}
it is given by
\begin{equation}
\label{retGreen}
  G_R(x,x')=i\vartheta(t-t')\big[\phi_0(x),\phi_0(x')\big],
\end{equation}
where the square brackets represent the commutator $\phi_0(x)\phi_0(x')
-\phi_0(x')\phi_0(x)$ and $\phi_0$ is the canonically quantized free field:
\begin{equation}
  \phi_0(x) = \int\dd\kk
  \left[\e^{i\kk\cdot\xx}\phi_k(t)\hat a_\kk + \mathrm{h.c}\right]
\end{equation}
where the modes $\phi_\kk(t)$ are the solution of the linear equation 
(obtained with $V=0$)
\begin{equation}
  \ddot\phi_k + 3H\dot\phi_k - \frac{k^2\phi_k}{a^2} + m^2\phi_k  = 0 \,.
\end{equation}

The objects we want to calculate are the \emph{in-in} correlation 
functions at equal times. As shown in \cite{Weinberg}, the expectation 
value between two \emph{in} vacua of a product of $r$ fields 
$Q(t)\equiv\phi(x_1)\dots\phi(x_r)$ all evaluated at the same time 
$t=t_1=\dots=t_r$ is given by
\begin{align}
\label{weinberg}
  \mean{Q(t)} &=\sum_{N=0}^\infty i^N
  \int_\tin^t \dd t_1 \int_\tin^{t_1} \dd t_2 \cdots 
  \int_\tin^{t_{N-1}} \dd t_N \notag \\
  &\times\mean{\Big[H_I(t_N),\cdots \!\Big[H_I(t_2),\! 
  \Big[H_I(t_1),Q_0(t)\Big]\Big]\Big]} \notag \\
  &\equiv \sum_{N=0}^\infty \mean{Q_N(t)};
\end{align}
in this equation, $Q_0(t)=\phi_0(x_1)\dots\phi_0(x_r)$ is the product 
of the same $m$ fields in the interaction picture, which was indicated
as $Q_I(t)$ in \cite{Weinberg}, and $H_I(t_i)$ is the interaction 
Hamiltonian.

Let us assume for simplicity that the interaction Hamiltonian is
\begin{equation}
  H_I(t_i) = \frac{\lambda}{n!}\int\dd \yy\,a^3(t_i)\phi_0^n(t_i,\yy);
\end{equation}
in this case, the $N$-th term of the series \eqref{weinberg} will be
the expectation value of
\begin{align}
\label{nestcomm}
  Q_N(t) &= 
  \left(\frac{i\lambda}{n!}\right)^N \!\! \int \prod_{i=1}^N
  \left[\dd y_i \,a^3(t_i) \,\vartheta(t_{i-1}-t_i)\right]\notag \\
  &\times\Big[\phi_0^n(y_N),\cdots\!\Big[\phi_0^n(y_2), 
  \!\Big[\phi_0^n(y_1),Q_0(t)\!\Big]\Big]\Big],
\end{align}
where we have defined $t_0\equiv t$ and $\dd y_i\equiv \dd t_i\dd\yy_i$.
In particular, choosing $Q(t)=\phi(x)$ we can write 
\begin{align}
\label{expansion}
  \phi(x) &=\sum_{N=0}^\infty
  \left(\frac{i\lambda}{n!}\right)^N \!\! \int \prod_{i=1}^N
  \left[\dd y_i \,a^3(t_i) \,\vartheta(t_{i-1}-t_i)\right]\notag \\
  &\times\Big[\phi_0^n(y_N),\cdots\!\Big[\phi_0^n(y_2), 
  \!\Big[\phi_0^n(y_1),\phi_0(x)\!\Big]\Big]\Big] \notag \\
  &\equiv \phi_0(x) + \sum_{N=1}^\infty\df_N(x).
\end{align}

Since each interaction is proportional to the coupling constant, the 
contribution of $N$-th order in $\lambda$, \ie $Q_N(t)$, contains $N$ 
nested commutators.
All these commutators involve multiple products of fields, and should be 
transformed into simpler commutators of two fields.
The reason for doing this is twofold: first, as shown in \cite{Weinberg}
commutators of two fields have a different late time behavior than fields
themselves, and it is useful to compute how many objects of each kind
there are in each contribution. Second, commutators of two fields
are just numbers commuting with everything, and can thus be pulled
out of any further commutator and expectation value.
After all the nested commutators are worked out, Eq.~\eqref{nestcomm} 
will split into the sum of several terms, each of which contains the 
product of $N$ two-field commutators times the expectation value of
the remaining free fields. This calculation can rapidly become rather
long and involved. Our goal is then to find a simple method to figure 
out which commutators and which expectation values of fields should 
appear in each term, with no need to go through all the commutator 
algebra.

We will now show how the $N$-th order term $Q_N(t)$ of the $r$-point
correlation function can be expressed as a combination of the $\df_i$'s 
appearing in the expansion \eqref{expansion} of the field.
Without loss of generality, we can set $r=2$ and consider the two-point 
correlation function $\mean{\phi(x_1)\phi(x_2)}$. The generalization 
to a higher number of fields is straightforward.

The 1-st order term contains the commutator
\begin{align}
  \left[\phi^n_0(y),\phi_0(x_1)\phi_0(x_2)\right] 
  &= \phi_0(x_1)\left[\phi^n_0(y),\phi_0(x_2)\right] \notag \\
  &+ \left[\phi^n_0(y),\phi_0(x_1)\right]\phi_0(x_2);
\end{align}
after being expressed in terms of commutators involving only one external 
field at time and plugged back into \eqref{nestcomm}, it yields
\begin{equation}
  Q_1(t) = \phi_0(x_1)\df_1(x_2) + \df_1(x_1)\phi_0(x_2).
\end{equation}
The second order term involves the commutator
\begin{align}
  \big[\phi_0^n(y_2),&\big[\phi_0^n(y_1),
  \phi_0(x_1)\phi_0(x_2)\big]\big] \notag \\
  &= \phi_0(x_1)\!\left[\phi_0^n(y_2),
  \left[\phi_0^n(y_1),\phi_0(x_2)\right]\right] \notag \\
  &+ \left[\phi_0^n(y_1),\phi_0(x_2)\right]\!
  \left[\phi_0^n(y_2),\phi_0(x_1)\right] \notag \\
  &+ \left[\phi_0^n(y_1),\phi_0(x_1)\right]\!
  \left[\phi_0^n(y_2),\phi_0(x_2)\right] \notag \\
  &+ \left[\phi_0^n(y_2),\!\left[\phi_0^n(y_1),\phi_0(x_1)\right]
  \right]\phi_0(x_2);
\end{align}
when inserted into \eqref{nestcomm}, the first and last terms immediately 
give $\phi_0(x_1)\df_2(x_2)$ and $\df_2(x_1)\phi_0(x_2)$ respectively. 
The integration variables $y_1$ and $y_2$ in the third term can be freely 
exchanged, which would make the second and third term equal. However,
the exchange also modifies the respective step functions, whose sum
now gives
\begin{gather}
  \vartheta(t-t_1)\vartheta(t_1-t_2)
  +\vartheta(t-t_2)\vartheta(t_2-t_1) \notag\\
  =\vartheta(t-t_1)\vartheta(t-t_2) \,;
\end{gather}
therefore, we obtain
\begin{align}
  Q_2(t) &= \phi_0(x_1)\df_2(x_2) + \df_2(x_1)\phi_0(x_2) \notag\\ 
  &+ \df_1(x_1)\df_1(x_2)\,.
\end{align}

This very same procedure can be iterated up to arbitrary order, yielding
\begin{equation}
\label{Q_N2p}
  Q_N(t) = \sum_{j=0}^N \df_j(x_1)\df_{N-j}(x_2),
\end{equation}
where the 0-th order terms are meant to be 
$\df_0(x_1)\equiv\phi_0(x_1)=\psi$ and $\df_0(x_2)\equiv\phi_0(x_2)=\chi$.
Summing up all perturbative orders one gets
\begin{equation}
  Q(t) = \sum_{N=0}^\infty Q_N(t) =
  \Bigg[\sum_{i=0}^\infty\df_i(x_1)\Bigg]
  \!\Bigg[\sum_{j=0}^\infty\df_j(x_2)\Bigg],
\end{equation}
which is what we could expect since we are considering 
$Q(t)=\phi(x_1)\phi(x_2)$.

Generalizing to the case of $r$ fields one has
\begin{equation}
\label{Q_Ngen}
  Q_N(t) = \sum_{|j|=N} \df_{j_1}(x_1)\cdots\df_{j_r}(x_r),
\end{equation}
where the sum is taken over all the vectors $j\in\mathbb{N}^r$ such that
$|j|\equiv j_1+\dots+j_r=N$. Again, this gives
\begin{equation}
  Q(t) = \sum_{N=0}^\infty Q_N(t)
  = \prod_{j=0}^r \Bigg[\sum_{i=0}^\infty\df_i(x_j)\Bigg].
\end{equation}

It is thus clear that, in order to reconstruct the expansion in powers
of the coupling constant $\lambda$ of any $r$-point correlation 
function $\mean{\phi(x_1)\cdots\phi(x_r)}$, one can just compute the
expansion \eqref{expansion}. Then one simply has to take the product
of all the possible combinations of the $\df_i$'s having the same 
overall number of powers of $\lambda$. Therefore, our goal will be
that of giving a simplified way to calculate \eqref{expansion}.


\section{Perturbative expansion of the equation of motion}
\label{pertexp}

We now want to find a simple method to calculate the evolution
of the interacting field as an expansion in powers of the coupling 
constant, in order to have a formulation equivalent to \eqref{expansion}
but still not relying on the involved computation of all the nested
commutators. In this Section we show that this can be achieved simply
with a perturbative solution of the equation of motion, once the
order of the non-commuting pieces of the expansion has been appropriately 
taken care of.

Let us then look at the same problem from a different perspective, and
solve perturbatively the equation of motion \eqref{eqmot} for the 
operator field
\begin{equation}
\label{pertsol}
  \phi(x)=\phi_0(x)+\df_1(x)+\df_2(x)+\dots\,,
\end{equation}
expanding around the solution with $\lambda=0$ (the field in the 
interaction picture).
Higher order contributions can be obtained by iteration, plugging
the known solution back into \eqref{eqmot}. The next order contribution
will then be the solution of a linear differential equation with a source.

The evolution equation for $\df_1$ (considering for simplicity 
a cubic potential $V=\frac{\lambda}{3!}\phi^3$) is
\begin{equation}
  \delta\ddot\phi_1 + 3H\delta\dot\phi_1 - \frac{\nabla^2\df_1}{a^2} 
  + m^2\df_1 = -\frac{\lambda}{2}\phi_0^2 \,,
\end{equation}
whose solution can be found with Green's method. 
Using the retarded Green function \eqref{retGreen}, one gets
\begin{equation}
\label{df1}
  \df_1(x) = -\frac{\lambda}{2}
  \int \dd y \,a_y^3 G_R(x,y) \phi_0^2(y) .
\end{equation}
When one wants to find the second order correction $\df_2$, 
the potential term in \eqref{eqmot} becomes $V'(\phi_0+\df_1)$ and 
must be expanded up to second order in $\lambda$. In this case
one must pay attention to the fact that $\phi_0$ and $\df_1$ do
not commute, since they are quantum operators and contain free fields
evaluated at different times. The expansion yields
\begin{equation}
  V'(\phi_0+\df_1) = \frac{\lambda}{2}\Big[\phi_0^2 +
  \phi_0\df_1 +\df_1\phi_0 + \mathcal{O}(\lambda^2)\Big];
\end{equation}
keeping only the second order terms one has the source for the 
equation of motion for $\df_2$, which reads
\begin{equation}
  \delta\ddot\phi_2 + 3H\delta\dot\phi_2 - \frac{\nabla^2\df_2}{a^2} 
  + m^2\df_2 = -\frac{\lambda}{2}\big\{\phi_0,\df_1\big\} \,,
\end{equation}
where the anticommutator 
$\big\{\phi_0,\df_1\big\} \equiv \phi_0\,\df_1+\df_1\,\phi_0$ 
symmetrizes the dependence on the non-commuting fields.
The solution can be found again with Green's method, and using the
expression \eqref{df1} for $\df_1$ gives
\begin{align}
  \df_2(x) &= \left(-\frac{\lambda}{2}\right)^2 \int \dd y \,a_y^3 
  \int \dd z \, a_z^3 \notag \\
  &\times G_R(x,y)G_R(y,z) \Big\{\phi_0(y),\phi_0^2(z)\Big\} ,
\label{df2}
\end{align}
Similarly, from the third order contribution to the potential one gets
\begin{align}
  \df_3 &= -\frac{\lambda}{2} \int \dd y \,a_y^3 G_{\!R}(x,y) 
  \Big[\!\big\{\phi_0(y),\df_2(y)\big\} + \df_1^2(y)\Big] \notag \\
  &= \left(-\frac{\lambda}{2}\right)^3 \int \dd y \,a_y^3
  \int \dd z \, a_z^3  \int \dd w \, a_w^3  G_R(x,y)\notag \\
  &\times\!\bigg[\Big\{\!\phi_0(y),G_R(y,z)
  \big\{\phi_0(z),G_R(z,w)\phi_0^2(w)\big\}\!\Big\} \notag \\
  &\quad + G_R(y,z)\phi_0^2(z)G_R(y,w)\phi_0^2(w)\bigg].
\label{df3}
\end{align}

More generally, including corrections to the field value up to $N$-th
order, the potential term in \eqref{eqmot} can be written as
\begin{equation}
  V'\!\left(\sum_{i=0}^N\df_i\right) =
  \frac{\lambda}{2}\sum_{i=0}^N\left(\df_i^2 +
  \sum_{j=i+1}^N\big\{\df_i,\df_j\big\}\right)\!,
\end{equation}
which allows to express iteratively the general solution as
\begin{subequations}
\label{dfn}
\begin{align}
  \df_{2i+1} &= - \frac{\lambda}{2} G_R \left[\df_i^2 +
  \sum_{j=0}^{i-1}\big\{\df_j,\df_{2i-j}\big\}\right]\!, \\
  \df_{2i+2} &= - \frac{\lambda}{2} G_R 
  \sum_{j=0}^i\big\{\df_j,\df_{2i+1-j}\big\}\,.
\end{align}
\end{subequations}

In general, any contribution to \eqref{expansion} is an integral of
a retarded Green function followed by two lower order $\df_i$'s. 
If these are not both $\phi_0$, the first $G_R$ is followed by another 
$G_R$, and so on. The generic contribution to $\df_N$ thus contains a 
tree of $N$ nested Green functions stemming from the first one. Each 
of them is followed by a pair of objects, which can be free fields or 
other Green functions. 
It also contains $N$ integrations over the $N$ vertices of the tree.
When the two objects after a Green function are not symmetric (like
a free field and a Green function, or two Green functions followed by
different objects), they must be symmetrized with an anticommutator. 
The total $\df_N$ is given by the sum of such contributions over all 
possible inequivalent dispositions $T_N$ of the $N$ Green functions.

When a generic interaction $\phi^n$ is considered, the previous
results still hold. The only difference is that any Green function
of the tree is now followed by $n-1$ objects. The symmetrization of 
asymmetric objects must now be done by mean of a generalized 
anticommutator. 
For $n-2$ (out of $n-1$) equal objects, this is defined as
\begin{align}
  \big\{\underbrace{\phi_1,\dots,\phi_1}_{n-2},\phi_2^{n-1}\big\}
  &\equiv \phi_1^{n-2}\phi_2^{n-1}
  + \phi_1^{n-3}\phi_2^{n-1}\phi_1 + \cdots \notag\\
  &+ \phi_1\phi_2^{n-1}\phi_1^{n-3} + \phi_2^{n-1}\phi_1^{n-2},
\label{genant}
\end{align}
which reduces to the standard anticommutator for $n=3$. In an 
arbitrary situation with $k_1$ free fields, $k_2$ groups of $n-1$ 
free fields and in general $k_i$ groups of $i(n-2)+1$ free fields 
(with $k_1+k_2+\dots=n-1$), this generalized anticommutator gives the 
$(n-1)!/(k_1!k_2!\cdots)$ inequivalent configurations.
The generic $\df_N$ then looks like
\begin{align}
  \df_N(x) &= \left(-\frac{\lambda}{2}\right)^N \sum_{T_N}
  \!\int \bigg[\prod_{i=1}^N \dd y_i \,a^3(t_i)\bigg] G_R(x,y_1) \notag \\
  &\times \big\{\dots ,G_R(y_{N-1},y_N)\phi_0^{n-1}(y_N)\big\},
\end{align}
where a number from 0 to $N-2$ of (generalized) anticommutators is 
needed depending on the number of vertices that has to be symmetrized.
The total number of free fields in each term of the sum is $N(n-2)+1$,
since each vertex of the tree one of them is replaced by a Green 
function carrying $n-1$ more free fields.

One can check that the iterative solution of the equation of motion gives 
the same results as the expansion \eqref{expansion}, after all the nested 
commutators have been reduced to products of commutators of two fields. 
In order to shorten the notation, we set $\phi_0(x)\equiv\psi$, 
and $\phi_0(y_i)\equiv\psi_i$ (not to be confused with the $\df_i$'s, for
which the index stands for the order of the expansion!).
The first order term in \eqref{expansion} contains the commutator 
$\left[\psi_1^3,\psi\right]=3\psi_1^2\left[\psi_1,\psi\right]$; recalling 
the definition \eqref{retGreen} of the Green function, the complete 
integrand reads
\begin{equation}
  \frac{i\lambda}{3!}\vartheta(t-t_1)\left[\phi_0^3(y_1),\phi_0(x)\right]=
  -\frac{\lambda}{2}G_R(x,y_1)\phi_0^2(y_1);
\end{equation}
once integrated, this is identical to the result \eqref{df1} obtained 
with the expansion of the equation of motion.
Similarly, the second order term of \eqref{expansion} contains
$ \left[\psi_2^3,\left[\psi_1^3,\psi\right]\right] =
3^2\big\{\psi_2^2,\psi_1\big\}\!
\left[\psi_2,\psi_1\right]\!\left[\psi_1,\psi\right]$,
and yields
\begin{align}
  &\left(\frac{i\lambda}{3!}\right)^{\!2}
  \vartheta(t-t_1)\vartheta(t_1-t_2)
  \Big[\phi_0^3(y_2),\left[\phi_0^3(y_1),\phi_0(x)\right]\!\Big] \notag\\
  &=\left(-\frac{\lambda}{2}\right)^{\!2}
  G_R(x,y_1)\Big\{\phi_0(y_1),G_R(y_1,y_2)\,\phi_0^2(y_2)\Big\};
\end{align}
here again each two-field commutator times the appropriate step function
has been turned into a Green function, returning the same expression  
as the integrand in \eqref{df2}.

If we want to go to the next order we need to add one more nested
commutator to those we just computed. The two-field commutators 
$\left[\psi_2,\psi_1\right]\!\left[\psi_1,\psi\right]$ in the inner
part are just numbers, and can be pulled out of the outer commutator. 
The latter then acts only on $\big\{\psi_2^2,\psi_1\big\}$. Since it 
obeys the Leibniz rule it will act separately on the two sides of the 
anticommutator, giving $\big\{\big[\psi_3^3,\psi_2^2\big],\psi_1\big\}
+ \big\{\psi_2^2,\big[\psi_3^3,\psi_1\big]\big\} =
3\big\{\big\{\psi_3^2,\psi_2\big\},\psi_1\big\}[\psi_3,\psi_2] +
3\big\{\psi_3^2,\psi_2^2\big\}[\psi_3,\psi_1]$.
Adding the step functions, the constants and the remaining two-field 
commutators, the first term becomes
\begin{gather}
  \left(-\frac{\lambda}{2}\right)^{\!3} \!
  G_R(x,y_1) G_R(y_1,y_2) G_R(y_2,y_3) \notag\\
  \times\Big\{\phi_0(y_1),\big\{\phi_0(y_2),\,\phi_0^2(y_3)\big\}\!\Big\},
\end{gather}
equalling the first term in \eqref{df3}.
In the second term, the $\vartheta$'s do not immediately match
the two-field commutators. However, $y_2$ and
$y_3$ are integration variables and can be exchanged, which gives
$\vartheta(t_1-t_2)\vartheta(t_2-t_3)\big\{\psi_3^2,\psi_2^2\big\}
= \vartheta(t_1-t_2) \vartheta(t_1-t_3) \psi_2^2\psi_3^2$; now the step 
functions both start at $t_1$, and we obtain
\begin{equation}
  \left(-\frac{\lambda}{2}\right)^{\!3}\!G_R(x,y_1)
  G_R(y_1,y_2)\,\phi_0^2(y_2)G_R(y_1,y_3)\,\phi_0^2(y_3)
\end{equation}
as in the second term of \eqref{df3}.

At each step, a higher order term can be obtained from a lower order one 
by turning one of its free fields into a new Green function. As shown,
this comes from the fact that when the new outer commutator acts on the 
free field it gives a two-field commutator times two new free fields.
One can show that the temporal step function can always be rearranged
so that the coordinates of one of them match those of the new two-field 
commutator in order to produce a Green function.
Finally, the Leibniz rule provide the necessary symmetrization.
All of the above can be applied as well to the case of a generic 
$\phi^n$ interaction. We therefore conclude that the perturbative 
expansion of the equation of motion for the operator field reproduces 
the very same results as the \emph{in-in} formalism in standard QFT.


\section{Graphical representation}
\label{graphrep}

As one can see, solving the equation of motion at some given order 
rapidly produces a very high number of terms, and also this procedure 
can thus become very complicated. Therefore, we would like to find
an automated method to reproduce efficiently all contributions to each 
term of \eqref{expansion}, and reconstruct their mathematical content.
In this Section, we introduce a graphical representation describing  
the perturbative expansion \eqref{expansion}. This will reach our goal,
allowing to construct perturbative corrections to the field evolution 
up to any order in $\lambda$ simply as a combination of graphs.

The field in the interaction picture, and the first and second order 
correction \eqref{df1} and \eqref{df2} can be represented respectively as
\begin{align}
  \phi_0(x) &=
\begin{pic}{35}{20}{(-5,-10)}
    \SetWidth{0.5}
    \SetColor{Black}
    \Text(0,5)[]{\normalsize{\Black{$x$}}}
    \Vertex(0,0){1.5}
    \DashLine(0,0)(20,0){3}
    \Cross(20,0)
\end{pic}, \\
\label{grphi_1}
  \df_1(x) &= 
\begin{pic}{50}{30}{(-5,-15)}
    \SetWidth{0.5}
    \SetColor{Black}
    \Text(0,5)[]{\normalsize{\Black{$x$}}}
    \Vertex(0,0){1.5}
    \ArrowLine(25,0)(0,0)
    \DashCArc(40,0)(15,90,270){3}
    \Text(22,4)[]{\normalsize{\Black{$y$}}}
    \Cross(40,15) \Cross(40,-15)
\end{pic}
\end{align}
and 
\begin{equation}
\label{grphi_2}
  \df_2(x)=
\begin{pic}{65}{60}{(-5,-30)}
    \SetWidth{0.5}
    \SetColor{Black}
    \Vertex(0,0){1.5}
    \ArrowLine(25,0)(0,0)
    \ArrowArc(45,0)(20,90,180)
    \DashCArc(45,0)(20,180,270){3}
    \Cross(45,-20)
    \DashCArc(55,20)(10,90,270){2}
    \Cross(55,30)\Cross(55,10)
    \Text(0,5)[]{\normalsize{\Black{$x$}}}
    \Text(22,4.5)[]{\normalsize{\Black{$y$}}}
    \Text(42,24)[]{\normalsize{\Black{$z$}}}
\end{pic}
  +
\begin{pic}{65}{60}{(-5,-30)}
    \SetWidth{0.5}
    \SetColor{Black}
    \Vertex(0,0){1.5}
    \ArrowLine(25,0)(0,0)
    \DashCArc(45,0)(20,90,180){3}
    \ArrowArcn(45,0)(20,270,180)
    \Cross(45,20)
    \DashCArc(55,-20)(10,90,270){2}
    \Cross(55,-30)\Cross(55,-10)
    \Text(0,5)[]{\normalsize{\Black{$x$}}}
    \Text(22,4.5)[]{\normalsize{\Black{$y$}}}
    \Text(41,-25)[]{\normalsize{\Black{$z$}}}
\end{pic}\;,
\end{equation}
where the rules to reassign to each element of the graph its 
original mathematical meaning are:
\begin{enumerate}
\item a dot is associated with the field coordinate $x$;
\item each solid line represents a retarded Green function, with the farther end from the dot in the past, and an arrow showing the flow of increasing time: 
\begin{equation*}
  \begin{pic}{28}{20}{(-5,-10)}
    \SetWidth{0.5} \SetColor{Black}
    \Vertex(0,0){1.5} \ArrowLine(20,0)(0,0)
    \Text(0,6)[]{\normalsize{\Black{$x$}}}
    \Text(21,5)[]{\normalsize{\Black{$y$}}}
  \end{pic}
  \,=\, G_R(x,y) \;,\quad
  \begin{pic}{28}{20}{(-5,-10)}
    \SetWidth{0.5} \SetColor{Black}
    \ArrowLine(20,0)(0,0)
    \Text(0,5)[]{\normalsize{\Black{$y$}}}
    \Text(22,5)[]{\normalsize{\Black{$z$}}}
  \end{pic}
  \,=\, G_R(y,z)\:;
\end{equation*}

\item each vertex carries an integration factor:
\begin{equation*}
  \begin{pic}{25}{20}{(-5,-10)}
      \SetWidth{0.5} \SetColor{Black}
      \Line(0,0)(13,0)\Line(13,0)(19,11)\Line(13,0)(19,-11)
    \Text(10,5)[]{\normalsize{\Black{$y_i$}}}
  \end{pic}
  \,=\, -\frac{\lambda}{(n-1)!}\int\dd y_i \,a^3(t_i)\:;
\end{equation*}

\item each dashed line with a cross is a free field, evaluated at the spacetime 
point to which it is attached:
\begin{equation*}
  \begin{pic}{25}{20}{(-5,-10)}
    \SetWidth{0.5} \SetColor{Black}
    \DashLine(0,0)(18,0){3}
    \Text(0,5)[]{\normalsize{\Black{$y$}}}
    \Line(16,-2)(20,2)\Line(16,2)(20,-2)
  \end{pic}
  \,=\, \phi_0(y)\:.
\end{equation*}
\end{enumerate}

For example, the interpretation of the graph \eqref{grphi_1} giving 
the first order correction $\delta\phi_1(x)$ is
\begin{equation}
\begin{pic}{115}{70}(-5,-35)
    \SetWidth{0.5}
    \SetColor{Black}
    \Vertex(0,0){2}
    \ArrowLine(45,0)(0,0)
    \DashCArc(70,0)(25,90,270){3}
    \Cross(70,25)\Cross(70,-25)
    \Text(90,25)[]{\normalsize{\Black{$\phi_0(y)$}}}
    \Text(90,-25)[]{\normalsize{\Black{$\phi_0(y)$}}}
    \Text(25,25)[]{\normalsize{\Black{%
            $-\displaystyle\frac{\lambda}{2}\int\mathrm{d} y\, a^3$}}}
    \LongArrow(33,18)(40,7)
    \Text(23,-8)[]{\normalsize{\Black{$G_R(x,y)$}}}
\end{pic}\,;
\end{equation}
here the two free fields are evaluated at the same time, and their 
overall contribution is simply $\phi_0^2(y)$, in the past of $\phi_0(x)$. 
This correctly reproduces the result of \eqref{df1}.

As for the second order contributions, the first of the two graphs 
in \eqref{grphi_2} gives
\begin{equation}
\begin{pic}{145}{80}(-20,-35)
    \SetWidth{0.5}
    \SetColor{Black}
    \Vertex(0,0){2}
    \ArrowLine(45,0)(0,0)
    \ArrowArc(70,0)(25,90,180)
    \DashCArc(70,0)(25,180,270){3}
    \DashCArc(80,25)(10,90,270){3}
    \Cross(70,-25)\Cross(80,35)\Cross(80,15)
    \Text(105,25)[]{\normalsize{\Black{$\bigg\}\;\phi_0^2(z)$}}}
    \Text(90,-25)[]{\normalsize{\Black{$\phi_0(y)$}}}
    \Text(10,35)[]{\normalsize{\Black{%
            $\displaystyle\left(\!-\frac{\lambda}{2}\right)^{\!2}
            \!\!\int\!\dd y \,a_y^3\dd z \,a_z^3$}}}
    \LongArrow(25,20)(40,7)
    \LongArrow(57,33)(66,28)
    \LongArrow(65,5)(56,15)
    \Text(23,-8)[]{\normalsize{\Black{$G_R(x,y)$}}}
    \Text(85,-3)[]{\normalsize{\Black{$G_R(y,z)$}}}
\end{pic}\,,
\end{equation}
where we get a double integration factor due to the presence of 
two vertices, and two retarded Green functions.
In this second order graph one also has to care about how to order 
the free fields, since these are now evaluated at unequal times and 
do not commute: we could get either $\phi_0^2(y_2)\phi_0(y_1)$ or 
$\phi_0(y_1)\phi_0^2(y_2)$.
Choosing to order the fields ``from top to bottom'' would give
\begin{equation}
\label{grphi_2_ord1}
\begin{pic}{140}{55}{(-5,-25)}
    \SetWidth{0.5}
    \SetColor{Black}
    \Vertex(0,0){1.5}
    \ArrowLine(25,0)(0,0)
    \ArrowArc(45,0)(20,90,180)
    \DashCArc(45,0)(20,180,270){3}
    \Cross(45,-20)
    \DashCArc(55,20)(10,90,270){2}
    \Cross(55,30)\Cross(55,10)
    \Text(0,5)[]{\normalsize{\Black{$x$}}}
    \Text(22,5)[]{\normalsize{\Black{$y$}}}
    \Text(40,24)[]{\normalsize{\Black{$z$}}}
    \Text(60,20)[l]{\normalsize{\Black{$\bigg\}$}}}
    \Text(66,4)[l]{\normalsize{\Black{$\left.\rule[-25pt]{0pt}{50pt}%
           \right\}\quad\phi_0^2(z)\phi_0(y)\;;$}}}
\end{pic}
\end{equation}
the second graph in \eqref{grphi_2} can be obtained from the previous 
one simply by flipping the lines in the first bifurcation, and thus 
differs only for the order of the fields. With the same convention
we would have
\begin{equation}
\label{grphi_2_ord2}
\begin{pic}{140}{50}{(-5,-30)}
    \SetWidth{0.5}
    \SetColor{Black}
    \Vertex(0,0){1.5}
    \ArrowLine(25,0)(0,0)
    \DashCArc(45,0)(20,90,180){3}
    \ArrowArcn(45,0)(20,270,180)
    \Cross(45,20)
    \DashCArc(55,-20)(10,90,270){2}
    \Cross(55,-30)\Cross(55,-10)
    \Text(0,5)[]{\normalsize{\Black{$x$}}}
    \Text(20,5)[]{\normalsize{\Black{$y$}}}
    \Text(40,-24)[]{\normalsize{\Black{$z$}}}
    \Text(60,-20)[l]{\normalsize{\Black{$\bigg\}$}}}
    \Text(66,-5)[l]{\normalsize{\Black{$\left.\rule[-25pt]{0pt}{50pt}%
             \right\}\quad\phi_0(y)\phi_0^2(z)\;,$}}}
\end{pic}
\end{equation}
and the sum of the two graphs
correctly reproduces $\df_2$ as given in \eqref{df2}.

Choosing a different convention to order the fields (for example,
``from bottom to top'') would reverse the contribution of each graph, 
but their sum would be unchanged. 
Therefore, independently of the rule chosen for the single graph, the
sum of the two mirroring graphs will always give the anticommutator.
We also notice that the second order graph contains two bifurcations,
but only one of them (the one labeled by $y$ in our graphs) introduces 
an anticommutator. The second bifurcation is symmetric (like the one in
the first order graph) and does not have any effect on the field order
when flipped. In general, only bifurcations with distinguishable branches
change the graph when they are flipped and introduce anticommutators.
It would be useful to denote the sum of such graphs as
\begin{equation}
\left\{
\begin{pic}{55}{50}{(-5,-20)}
    \SetWidth{0.5}
    \SetColor{Black}
    \Vertex(0,0){1.5}
    \ArrowLine(20,0)(0,0)
    \ArrowArc(35,0)(15,90,180)
    \DashCArc(35,0)(15,180,270){3}
    \Cross(35,-15)
    \DashCArc(45,15)(10,90,270){2}
    \Cross(45,25)\Cross(45,5)
    \Text(0,5)[]{\normalsize{\Black{$x$}}}
    \Text(17,5)[]{\normalsize{\Black{$y$}}}
    \Text(32,19)[]{\normalsize{\Black{$z$}}}
\end{pic}
\right\}_{\!2} ,
\end{equation}
where the index stands for the number of graphs produced by the
anticommuting bifurcation. Independently of the convention chosen
to order the free fields, this symmetrized graph directly yields
the anticommutator $\{\phi_0(y),\phi_0^2(z)\}$.

In general, each correction $\df_N(x)$ is represented by a tree-shaped 
graph with $N$ bifurcations connected by lines representing the retarded 
Green functions. Higher bifurcations therefore correspond to points in 
the past light cone of the space-time coordinate $x$ and of lower
bifurcations. The final branches of each tree graph represents all
the free fields contained in the expansion, of which we will take the
expectation value. 
The previous examples suggest that asymmetric graphs should always be 
considered together with its mirror counterpart(s): the sum of all the
graphs lead to the correct symmetrization procedure of these
non-commuting free fields. More precisely, 
one has to consider the sum of all the graphs that can be obtained from 
each other simply by flipping the branches of one or more bifurcations.
the sum of all graphs will then contain as many anticommutators 
as distinguishable bifurcations, with each anticommutator involving the 
fields (or groups of fields) attached to the branches that are being 
flipped.

Following the same steps we can write the third order expansion
\eqref{df3} as
\begin{equation}
\label{grphi_3}
  \df_3(x)=
\begin{pic}{55}{50}{(-5,-25)}
    \SetWidth{0.5}
    \SetColor{Black}
    \Vertex(0,0){1.5}
    \ArrowLine(20,0)(0,0)
    \ArrowArc(35,0)(15,90,180)
    \ArrowArcn(35,0)(15,270,180)
    \DashCArc(45,15)(10,90,270){3}
    \DashCArc(45,-15)(10,90,270){3}
    \Cross(45,25)\Cross(45,5)\Cross(45,-5)\Cross(45,-25)
    \Text(0,5)[]{\normalsize{\Black{$x$}}}
    \Text(15,5)[]{\normalsize{\Black{$y$}}}
    \Text(30,19)[]{\normalsize{\Black{$z$}}}
    \Text(30,-19)[]{\normalsize{\Black{$w$}}}
\end{pic}
  + 
\left\{\!
\begin{pic}{60}{50}{(-5,-25)}
    \SetWidth{0.5}
    \SetColor{Black}
    \Vertex(0,0){1.5}
    \ArrowLine(20,0)(0,0)
    \ArrowArc(35,0)(15,90,180)
    \DashCArc(35,0)(15,180,270){3}
    \DashCArc(45,15)(10,90,180){2}
    \ArrowArcn(45,15)(10,270,180)
    \DashCArc(52,5)(7,90,270){2}
    \Cross(35,-15)
    \Cross(45,25)
    \Cross(52,12)\Cross(52,-2)
    \Text(0,5)[]{\normalsize{\Black{$x$}}}
    \Text(15,5)[]{\normalsize{\Black{$y$}}}
    \Text(30,19)[]{\normalsize{\Black{$z$}}}
    \Text(40,0)[]{\normalsize{\Black{$w$}}}
\end{pic}
\right\}_{\!2,2}\;,
\end{equation}
where the double index in the second graph means that we now have 
two bifurcations ($y$ and $z$) with two possibilities each, for a 
total of four graphs that we must sum. 
Flipping the bifurcation in $z$ exchanges $\phi_0(z)$ with 
$\phi_0^2(w)$, while flipping the first one exchanges $\phi_0(y)$ 
with all the other fields taken together.
This will therefore results in the double nested anticommutator
$\{\phi_0(y),\{\phi_0(z),\phi_0^2(w)\}\}$, as prescribed by
\eqref{df3}. On the other side the first graph does not contain any
distinguishable bifurcation: indeed, its free field content is globally 
symmetric because, although $\phi_0^2(z)$ and $\phi_0^2(w)$ do not 
commute, $z$ and $w$ are dummy integration variables and can be safely 
exchanged without affecting the overall result.

When dealing with a generic interaction such as $\phi^n$, each vertex
will contain $n-1$ outgoing branches. If only one branch has one more 
bifurcation attached, then there will be $n-1$ possibilities of twisting 
the ``$(n-1)$-furcation''. If the branches containing further pieces are 
two, we will have $(n-1)(n-2)$ possibilities if the sub-graphs of the two 
branches are distinguishable and $(n-1)(n-2)/2$ if they are not.
In general, the number of possible dispositions of $k$ equal objects 
in $n-1$ slots is given by the binomial coefficient $\binom{n-1}{k}$.
In this case, we need to use the generalized anticommutator
\eqref{genant}.

We can thus complete our set of prescriptions with the following 
rule, which describes the symmetrization procedure of the free fields
in asymmetric graphs by mean of anticommutators:
\begin{itemize}
\item[5.] the sum of all the mirroring graphs with $l$ distinguishable 
bifurcations contains $l$ (generalized) anticommutators exchanging the 
group of fields attached to each branch of the bifurcation.
\end{itemize}

We can easily show that this graphical method is correct up to any 
order. First of all, we notice that the second order graphs in
\eqref{grphi_2} can be obtained combining together two first order
graphs in the two possible positions. Also the third order graphs
in \eqref{grphi_3} are obtained attaching two first order or one second 
order graph respectively on the branches of a first order one. 
This reflects the structure of the analytic expressions \eqref{df1}
and \eqref{df2} for $\df_1$ and $\df_2$, where there is an overall
Green function $G_R$. According to our graphic rules, this is actually
represented by the ``handle'' of the first bifurcation, to which
one has to attach other lower order pieces.
Generalizing, since as shown in \eqref{dfn} all corrections have an 
overall Green function $G_R$, then the graphical procedure of obtaining 
higher order corrections from combining lower order terms on the top of 
a first order graph holds at any order.

However, constructing $N$-th order graphs is actually much simpler, 
since one can simply write down all the possible graphs with $N$ 
bifurcations in all possible positions, without caring for 
lower order graphs. The equivalence of the two procedures is
implicit in the recursive structure of Eq.~\eqref{dfn} for a generic 
$\df_N$. The proof is very simple: Eq.~\eqref{dfn} shows that the
$N$-th order is made up by all the possible combinations of lower
order graphs containing a total amount of bifurcations equal to $N-1$,
attached on the top of a first order graph. The lower order sub-graphs 
are disposed in all possible positions, as a consequence of the presence 
of the anticommutators.
Therefore, if the two procedure are equivalent up to the $(N-1)$-th 
order, then they are also up the $N$-th.

So far, we have been describing the graphical method to write
down \emph{in-in} expectation values in the case of a cubic interaction,
where one only has two-ways bifurcations. However, the procedure is
completely general, and can be used to describe a $\phi^n$ theory
where ramifications with $n-1$ branches exist. The steps one needs to
follow are strictly the same; the only difference is that there
are now several different ways to flip the branches of each
$(n-1)$-furcation, which is a graphical representation of the generalized
anticommutator \eqref{genant}. For example, the second order
contribution in a $\lambda\phi^4$ theory is
\begin{equation}
  \df_2(x)= \left\{\!
\begin{pic}{65}{60}{(-5,-30)}
    \SetWidth{0.5}
    \SetColor{Black}
    \Vertex(0,0){1.5}
    \ArrowLine(25,0)(0,0)
    \ArrowArc(45,0)(20,90,180)
    \DashCArc(45,0)(20,180,270){3}
    \DashLine(25,0)(45,0){3}
    \Cross(45,-20) \Cross(45,0)
    \DashCArc(55,20)(10,90,270){2}
    \DashLine(45,20)(55,20){2}
    \Cross(55,30)\Cross(55,20)\Cross(55,10)
    \Text(0,5)[]{\normalsize{\Black{$x$}}}
    \Text(22,5)[]{\normalsize{\Black{$y$}}}
    \Text(42,24)[]{\normalsize{\Black{$z$}}}
\end{pic}\right\}_{\!3},
\end{equation}
where the sum over the three possible graphs obtained from the 
permutation of the branches produces the generalized anticommutator
\begin{equation}
  \phi_0^3(z)\phi_0^2(y) + \phi_0(y)\phi_0^3(z)\phi_0(y)
  + \phi_0^2(y)\phi_0^3(z).
\end{equation}

Summarizing, the rules for constructing an $N$-th order graph with a 
generic interaction $\phi^n$ are
\begin{itemize}
\item start the perturbative tree from the external point, connecting 
any additional vertex to lower vertices or to the external point with 
a solid line
\item each vertex has one ingoing solid line (connecting to the lower 
part of the tree), and $n-1$ (solid or dashed) outgoing lines
\item outgoing solid lines lead to higher vertices, outgoing dashed
lines are final. When connecting an additional vertex, a dashed line
must be changed into a solid one.
\item the total number of vertices must equal the order of the graph 
\end{itemize}
One must then take the sum over all the possible disposition of the $N$
vertices, and apply the previous rules 1 to 5 in order to interpret 
the graphs.


\section{Feynman rules for correlation functions}
\label{Feyn}

In this Section, we combine several graphical perturbative solutions 
in order to recover closed Feynman-like diagrams for the correlation 
functions. Closed graphs will therefore represent the expectation
value of the free field content described by some combination of open 
tree-graphs.

As we saw in \eqref{Q_N2p} and \eqref{Q_Ngen}, the product $Q(t)$
of some interacting fields can be expanded into the sum of terms 
involving the product of several $\df_i$'s. For example, the
two-point function can be written as
\begin{equation}
  \mean{\phi(x_1)\phi(x_2)}=\sum_{N=0}^\infty
  \sum_{j=0}^N \mean{\df_j(x_1)\df_{N-j}(x_2)},
\end{equation}
where each expectation value acts on the product of all the Gaussian 
free fields contained in the $\df_i$'s. Using Wick's theorem, all these 
expectation values can be in turn decomposed into several products of 
two-field expectation values, containing as usual all the possible 
pairs of the free fields.

We now want to implement this fact into the graphical representation.
From a graphical point of view, an $m$-point correlation function can 
thus be reproduced by putting together $m$ field expansion diagrams 
(each of which is associated to an external point) and connecting
all their free fields in all possible ways.
We will graphically describe the contraction of two free fields as
\begin{equation}
  \mean{
  \begin{pic}{55}{14}{(-5,-7)}
    \SetWidth{0.5}
    \SetColor{Black}
    \Text(0,4)[]{\normalsize{\Black{$y$}}}
    \DashLine(0,0)(20,0){3}
    \Cross(20,0)\Cross(25,0)
    \DashLine(25,0)(45,0){3}
    \Text(45,5)[]{\normalsize{\Black{$z$}}}
  \end{pic}
  } \equiv
  \begin{pic}{40}{14}{(-5,-7)}
    \SetWidth{0.5}
    \SetColor{Black}
    \Text(0,4)[]{\normalsize{\Black{$y$}}}
    \DashLine(0,0)(30,0){3}\Cross(15,0)
    \Text(30,5)[]{\normalsize{\Black{$z$}}}
  \end{pic},
\end{equation}
which as we will see can represent both the expectation values
$\mean{\phi_0(y)\phi_0(z)}$ or $\mean{\phi_0(z)\phi_0(y)}$.
We are going to clarify this point with some examples.

The simplest case to study is the mean value of the field.
We immediately see that all the even terms of the expansion vanish, 
since they contain an odd number of free fields. We thus have
$\mean{\df_N(x)}=0$ for $N$ even. For the first order term, we have 
instead 
\begin{equation}
\label{gr1p}
  \mean{\df_1(x)} =
  \Big\langle
  \begin{pic}{41}{30}{(-5,-15)}
    \SetWidth{0.5}
    \SetColor{Black}
    \Text(0,5)[]{\normalsize{\Black{$x$}}}
    \Vertex(0,0){1.5}
    \ArrowLine(20,0)(0,0)
    \Text(18,5)[]{\normalsize{\Black{$y$}}}
    \DashCArc(30,0)(10,90,270){3}
    \Cross(30,10)\Cross(30,-10)
    \Line(33,10)(35,10)\DashLine(35,10)(35,-10){2}\Line(33,-10)(35,-10)
  \end{pic} \;\Big\rangle =
  \begin{pic}{50}{30}{(-5,-15)}
    \SetWidth{0.5}
    \SetColor{Black}
    \Vertex(0,0){1.5}
    \Text(0,5)[]{\normalsize{\Black{$x$}}}
    \ArrowLine(20,0)(0,0)
    \Text(18,5)[]{\normalsize{\Black{$y$}}}
    \DashCArc(30,0)(10,0,360){3}
    \Cross(40,0)
  \end{pic},
\end{equation}
where the dashed line represents the only available contraction of the
two free fields, which is $\mean{\phi^2(y)}$, while the meaning of
all the other graphic elements (solid lines, vertices, points) remains 
unchanged.
Since there is only one possibility to contract the free fields, the 
overall factor of $\frac{1}{2}$ coming from the vertex remains.
The result is therefore
\begin{equation}
  \mean{\df_1(x)} = -\frac{\lambda}{2}
  \int \dd y \,a^3 G_R(x,y) \!\mean{\phi_0^2(y)} .  
\end{equation}

Another useful example is the three-point function 
$\mean{\phi(x_1)\phi(x_2)\phi(x_3)}$. At lowest (that is, first) 
order in $\lambda$, this is given by the contraction of one first order 
and two zeroth order terms:
\begin{align}
  \mean{\phi(x_1)\phi(x_2)\phi(x_3)} &\simeq
  \mean{\df_1(x_1)\phi_0(x_2)\phi_0(x_3)} \notag \\
  &+ \mean{\phi_0(x_1)\df_1(x_2)\phi_0(x_3)} \notag \\
  &+ \mean{\phi_0(x_1)\phi_0(x_2)\df_1(x_3)},
\label{3p}
\end{align}
while all the even terms of the expansion vanish.

The first term of \eqref{3p} contains the Green function $G_R(x_1,y)$ and 
a total free field content given by $\phi_0^2(y)\phi_0(x_2)\phi_0(x_3)$, 
whose expectation value is
\begin{align}
  \mean{\phi_0^2(y)\phi_0(x_2)\phi_0(x_3)} &=
  2\mean{\phi_0(y)\phi_0(x_2)}\!\mean{\phi_0(y)\phi_0(x_3)} \notag \\
  &+ \mean{\phi_0^2(y)}\!\mean{\phi_0(x_2)\phi_0(x_3)}.
\label{expect}
\end{align}
The last part of this expression would merely
give rise to the disconnected contribution 
$\mean{\df_1(x_1)}\!\mean{\phi_0(x_2)\phi_0(x_3)}$, which we will 
neglect.
The connected part of the first term of \eqref{3p} can thus be 
represented as
\begin{equation}
  \bigg\langle
\begin{pic}{70}{30}{(-5,-15)}
  \SetWidth{0.5}
  \SetColor{Black}
  \Text(1,5)[]{\normalsize{\Black{$x_1$}}}
  \Vertex(0,0){1.5}
  \ArrowLine(20,0)(0,0)
  \Text(17,4.5)[]{\normalsize{\Black{$y$}}}
  \DashCArc(30,0)(10,90,270){3}
  \Cross(30,10)\Cross(30,-10)
  \Cross(35,10) \DashLine(35,10)(50,10){2}
  \Vertex(50,10){1.5}
  \Text(60,10)[]{\normalsize{\Black{$x_2$}}}
  \Cross(35,-10) \DashLine(35,-10)(50,-10){2}
  \Vertex(50,-10){1.5}
  \Text(60,-11)[]{\normalsize{\Black{$x_3$}}}
\end{pic} \;\bigg\rangle =
\begin{pic}{50}{40}{(-5,-20)}
  \SetWidth{0.5}
  \SetColor{Black}
  \Vertex(0,0){1.3}
  \Text(2,5)[]{\normalsize{\Black{$x_1$}}}
  \ArrowLine(20,0)(0,0)
  \Text(18.5,4)[]{\normalsize{\Black{$y$}}}
  \DashLine(20,0)(30,17){3}\DashLine(20,0)(30,-17){3}
  \Text(25,9)[]{\rotatebox{15}{\scriptsize{\textbf{+}}}}
  \Text(25,-9)[]{\rotatebox{75}{\scriptsize{\textbf{+}}}}
  \Vertex(30,17){1.3}\Vertex(30,-17){1.3}
  \Text(40,17)[]{\normalsize{\Black{$x_2$}}}
  \Text(40,-18)[]{\normalsize{\Black{$x_3$}}}
\end{pic},
\end{equation} 
where the solid line represents as before the retarded Green function 
$G_R(x_1,y)$ ($y$ being the coordinate of the vertex), while the dashed 
lines stand for the contractions $\mean{\phi_0(y)\phi_0(x_2)}$ and
$\mean{\phi_0(y)\phi_0(x_3)}$. 
Unlike in \eqref{gr1p}, the free fields are now evaluated at unequal 
times and their order is relevant. 
In particular, the free field contribution $\phi_0^2(y)$ contained in 
$\df_1(x_1)$ will always appear before $\phi_0(x_2)$ or $\phi_0(x_3)$ 
in any expectation value.

This has to be repeated for the other two possible configurations in
\eqref{3p}. The complete representation of the first order contribution 
to the three-point function \eqref{3p} therefore becomes
\begin{equation}
\label{gr3p}
\begin{pic}{50}{40}{(-5,-20)}
  \SetWidth{0.5}
  \SetColor{Black}
  \Vertex(0,0){1.3}
  \Text(2,5)[]{\normalsize{\Black{$x_1$}}}
  \ArrowLine(20,0)(0,0)
  \DashLine(20,0)(30,17){3}\DashLine(20,0)(30,-17){3}
  \Text(25,9)[]{\rotatebox{15}{\scriptsize{\textbf{+}}}}
  \Text(25,-9)[]{\rotatebox{75}{\scriptsize{\textbf{+}}}}
  \Vertex(30,17){1.3}\Vertex(30,-17){1.3}
  \Text(40,17)[]{\normalsize{\Black{$x_2$}}}
  \Text(40,-18)[]{\normalsize{\Black{$x_3$}}}
\end{pic}\!\!\!+
\begin{pic}{50}{40}{(-5,-20)}
  \SetWidth{0.5}
  \SetColor{Black}
  \Vertex(0,0){1.3}
  \Text(2,5)[]{\normalsize{\Black{$x_1$}}}
  \DashLine(0,0)(20,0){3} \Cross(10,0)
  \ArrowLine(20,0)(30,17)\DashLine(20,0)(30,-17){3}
  \Text(25,-9)[]{\rotatebox{75}{\scriptsize{\textbf{+}}}}
  \Vertex(30,17){1.3}\Vertex(30,-17){1.3}
  \Text(40,17)[]{\normalsize{\Black{$x_2$}}}
  \Text(40,-18)[]{\normalsize{\Black{$x_3$}}}
\end{pic}\!\!\!+
\begin{pic}{50}{40}{(-5,-20)}
  \SetWidth{0.5}
  \SetColor{Black}
  \Vertex(0,0){1.3}
  \Text(2,5)[]{\normalsize{\Black{$x_1$}}}
  \DashLine(0,0)(20,0){3} \Cross(10,0)
  \DashLine(20,0)(30,17){3} \ArrowLine(20,0)(30,-17)
  \Text(25,9)[]{\rotatebox{15}{\scriptsize{\textbf{+}}}}
  \Vertex(30,17){1.3}\Vertex(30,-17){1.3}
  \Text(40,17)[]{\normalsize{\Black{$x_2$}}}
  \Text(40,-18)[]{\normalsize{\Black{$x_3$}}}
\end{pic},
\end{equation}
where in each graph the solid line is attached to a different external 
point. Having said about the first graph, in the remaining two the
solid lines represent the Green functions $G_R(x_2,y)$ and $G_R(x_3,y)$ 
respectively. 
As for the dashed lines, the free field content of the last two graphs is 
$\phi_0(x_1)\phi_0^2(y)\phi_0(x_3)$ and $\phi_0(x_1)\phi_0(x_2)\phi_0^2(y)$:
therefore, the dashed lines stand for $\mean{\phi_0(x_1)\phi_0(y)}$ and 
$\mean{\phi_0(y)\phi_0(x_3)}$ in the second graph, and finally for 
$\mean{\phi_0(x_1)\phi_0(y)}$ and $\mean{\phi_0(x_2)\phi_0(y)}$ in the 
third.

From this example we can extrapolate the rule for the interpretation
of such ``external'' dashed lines (\ie connecting different perturbative 
expansions evaluated at different spatial points):  
the order of the fields in ``external'' contractions is the same as
the respective position in the correlation function of the perturbative 
expansions they belong to.
Following this rule, the diagrammatic representation \eqref{gr3p} of
the three-point function correctly gives
\begin{align}
  &\mean{\phi(x_1)\phi(x_2)\phi(x_3)} \simeq \lambda\int\dd y \,a^3 \notag \\
  & \times\Big[G_R(x_1,y)
  \!\mean{\phi_0(y)\phi_0(x_2)}\!\mean{\phi_0(y)\phi_0(x_3)} \notag \\
  &
  + \mean{\phi_0(x_1)\phi_0(y)}\!G_R(x_2,y)\!\mean{\phi_0(y)\phi_0(x_3)}
  \notag \\
  &
  + \mean{\phi_0(x_1)\phi_0(y)}\!\mean{\phi_0(x_2)\phi_0(y)}\!G_R(x_3,y)\Big]
\end{align}
up to terms of order $\mathcal{O}(\lambda^3)$.


It is also instructive to examine the two-point correlation function.
For example, combining the two first order contributions $\df_1(x_1)$
and $\df_1(x_2)$ (which give a second order contribution) we have
\begin{equation}
\label{gr2p}
  \bigg\langle
  \begin{pic}{75}{30}{(-5,-15)}
    \SetWidth{0.5}
    \SetColor{Black}
    \Text(0,5)[]{\normalsize{\Black{$x_1$}}}
    \Vertex(0,0){1.5}
    \ArrowLine(20,0)(0,0)
    \DashCArc(30,0)(10,90,270){3}
    \Cross(30,10)\Cross(30,-10)
    \Cross(35,10)\Cross(35,-10)
    \DashCArc(35,0)(10,270,90){3}
    \ArrowLine(45,0)(65,0)
    \Vertex(65,0){1.5}
    \Text(17,4)[]{\normalsize{\Black{$y$}}}
    \Text(48,4)[]{\normalsize{\Black{$z$}}}
    \Text(65,5)[]{\normalsize{\Black{$x_2$}}}
  \end{pic} \bigg\rangle =
  \begin{pic}{60}{30}{(-5,-15)}
    \SetWidth{0.5}
    \SetColor{Black}
    \Vertex(0,0){1.5}
    \Text(0,5)[]{\normalsize{\Black{$x_1$}}}
    \ArrowLine(15,0)(0,0)
    \DashCArc(25,0)(10,0,360){3}
    \Cross(25,10)
    \Cross(25,-10)
    \ArrowLine(35,0)(50,0)
    \Vertex(50,0){1.5}
    \Text(12,-5)[]{\normalsize{\Black{$y$}}}
    \Text(38,-5)[]{\normalsize{\Black{$z$}}}
    \Text(50,5)[]{\normalsize{\Black{$x_2$}}}
  \end{pic},
\end{equation}
where the solid lines give the two Green functions $G_R(x_1,y)$ and 
$G_R(x_2,z)$, and the dashed lines represent the double expectation 
value $\mean{\phi_0(y)\phi_0(z)}^2$. Also in this case, the dashed
lines are of the ``external'' kind, connecting free fields in
different perturbative expansions.
The order of the fields in the expectation value is thus fixed by their
position in the correlation function $\mean{\phi(x_1)\phi(x_2)}$, since 
$\phi_0(y)$ belongs to the expansion of $\phi(x_1)$ and $\phi_0(z)$ 
to that of $\phi(x_2)$.
Finally, each vertex brings a factor of $\frac{1}{2}$, and since there 
are two equivalent ways to pair the fields this yields an overall 
symmetry factor of $\frac{1}{2}$.

Another second order contribution to the two-point function is obtained
combining $\df_2(x_1)$ with $\phi_0(x_2)$. From the two graphs in
Eq.~\eqref{grphi_2} one gets
\begin{align}
\label{pair2p}
\left\langle
\begin{pic}{80}{30}{(-5,-15)}
    \SetWidth{0.5}
    \SetColor{Black}
    \Vertex(0,0){1.5}
    \ArrowLine(20,0)(0,0)
    \ArrowArc(30,0)(10,90,180)
    \DashCArc(30,0)(10,180,270){3}
    \Cross(30,-10)
    \DashCArc(37,10)(7,90,270){3}
    \Cross(37,17)\Cross(37,3)
    \DashLine(40,3)(42,3){2}\DashLine(42,3)(42,-10){2}
    \DashLine(34,-10)(42,-10){2}
    \Text(1,5)[]{\normalsize{\Black{$x_1$}}}
    \DashLine(40,17)(50,17){2}\DashLine(50,17)(50,0){2}
    \DashLine(50,0)(52,0){2}
    \Cross(55,0) \DashLine(55,0)(70,0){2}
    \Vertex(70,0){1.5}
  \Text(17,4)[]{\normalsize{\Black{$y$}}}
  \Text(27,14)[]{\normalsize{\Black{$z$}}}
  \Text(71,5)[]{\normalsize{\Black{$x_2$}}}
  \end{pic}
\right\rangle +
\left\langle
\begin{pic}{80}{30}{(-5,-15)}
    \SetWidth{0.5}
    \SetColor{Black}
    \Vertex(0,0){1.5}
    \ArrowLine(20,0)(0,0)
    \ArrowArcn(30,0)(10,270,180)
    \DashCArc(30,0)(10,90,180){3}
    \Cross(30,10)
    \DashCArc(37,-10)(7,90,270){3}
    \Cross(37,-17)\Cross(37,-3)
    \DashLine(40,-3)(42,-3){2}\DashLine(42,-3)(42,10){2}
    \DashLine(34,10)(42,10){2}
    \Text(1,5)[]{\normalsize{\Black{$x_1$}}}
    \DashLine(40,-17)(50,-17){2}\DashLine(50,-17)(50,0){2}
    \DashLine(50,0)(52,0){2}
    \Cross(55,0) \DashLine(55,0)(70,0){2}
    \Vertex(70,0){1.5}
    \Text(17,4)[]{\normalsize{\Black{$y$}}}
    \Text(27,-13)[]{\normalsize{\Black{$z$}}}
    \Text(71,5)[]{\normalsize{\Black{$x_2$}}}
  \end{pic}
\right\rangle 
\end{align}
where in both these graphs the solid line contribution reads
$G_R(x_1,y)G_R(y,z)$.
The line attached to $x_2$ is again an external dashed line:
since in both cases all the free fields in $\df_2(x_1)$ precedes 
$\phi_0(x_2)$, this represents in both graphs\footnote{We are 
not considering here the possible pairing of $\phi_0(x_2)$ with 
$\phi_0(y)$, since this would not give a 1PI diagram} the expectation
value $\mean{\phi_0(z)\phi_0(x_2)}$.
On the other hand, the two free fields $\phi_0(y)$ and $\phi_0(z)$ both 
belong to $\df_2(x_1)$ and when we pair them we must pay attention to 
their order.
If we use the convention ``from top to bottom'' previously applied 
in \eqref{grphi_2_ord1} and \eqref{grphi_2_ord2}, then this
dashed line reads $\mean{\phi_0(z)\phi_0(y)}$ in the first graph
and $\mean{\phi_0(y)\phi_0(z)}$ in the second one. 
As for the diagrammatic representation of $\df_2$, using a different
convention would reverse the order of the fields in the internal dashed
line of each diagram; still, their sum would be unchanged.

It will thus prove useful to use a single diagram to denote the sum
of the two contributions in \eqref{pair2p}. Deforming the lines one
gets
\begin{equation}
\label{gr2p_2}
  \begin{pic}{62}{30}{(-5,-15)}
    \SetWidth{0.5}
    \SetColor{Black}
    \Vertex(0,0){1.5}
    \ArrowLine(15,0)(0,0)
    \ArrowArc(25,0)(10,0,180)
    \DashCArc(25,0)(10,180,0){3}\Cross(25,-10)
    \DashLine(35,0)(50,0){3}\Cross(43,0)
    \Vertex(50,0){1.5}
    \Text(2,5)[]{\normalsize{\Black{$x_1$}}}
    \Text(12,-4)[]{\normalsize{\Black{$y$}}}
    \Text(37,-4)[]{\normalsize{\Black{$z$}}}
    \Text(52,5)[]{\normalsize{\Black{$x_2$}}}
  \end{pic} \quad\left[ =
  \begin{pic}{65}{34}{(-5,-17)}
    \SetWidth{0.5}
    \SetColor{Black}
    \Vertex(0,0){1.5}
    \ArrowLine(15,0)(0,0)
    \ArrowArcn(25,0)(10,0,180)
    \DashCArc(25,0)(10,0,180){3}\Cross(25,10)
    \DashLine(35,0)(50,0){3}\Cross(43,0)
    \Vertex(50,0){1.5}
    \Text(2,5)[]{\normalsize{\Black{$x_1$}}}
    \Text(12,-4)[]{\normalsize{\Black{$y$}}}
    \Text(37,-4)[]{\normalsize{\Black{$z$}}}
    \Text(52,5)[]{\normalsize{\Black{$x_2$}}}
  \end{pic}\right]
\end{equation}
(the two graphs are completely equivalent, since they both represent
the sum of the two terms).
We now have here two different kind of dashed lines: again an external 
line (the one attached to $x_2$) connecting two different perturbative 
expansions, but also a dashed line between points that are also 
connected by solid lines (\ie belonging to the same perturbative 
expansion). In the language of closed graphs, this can be dubbed an 
``internal'' dashed line. 
The two dashed lines represent the expectation value
\begin{align}
  &\mean{\big\{\phi_0(y),\phi_0^2(z)\big\}\phi_0(x_2)} =
  2\mean{\phi_0(z)\phi_0(x_2)} \notag \\[-5pt]
&\qquad
\begin{pic}{70}{0}{(-5,0)}
  \Line(0,0)(0,3)\DashLine(0,0)(26,0){3}\Line(26,0)(26,3)
  \Line(29,0)(29,3)\DashLine(29,0)(55,0){3}\Line(55,0)(55,3)
\end{pic} \notag\\
&\qquad\qquad\qquad
\times\big[\mean{\phi_0(y)\phi_0(z)}+\mean{\phi_0(z)\phi_0(y)}\big],
\end{align}
where the free fields paired through the external dashed line are 
``pulled out'' of the anticommutator, while those paired with the 
internal dashed line are symmetrized.

Following the same procedure we can calculate the last remaining 
second order term, which combines $\phi_0(x_1)$ with the two graphs for
$\df_2(x_2)$. This gives
\begin{equation}
\label{gr2p_3}
\left\langle
\begin{pic}{92}{30}{(-5,-15)}
    \SetWidth{0.5}
    \SetColor{Black}
    \Vertex(0,0){1.5}
    \DashLine(0,0)(13,0){3}\Cross(13,0)
    \Text(2,5)[]{\normalsize{\Black{$x_1$}}}
    \DashLine(16,0)(27,0){2}
    \DashLine(27,0)(27,17){2}\DashLine(27,17)(35,17){2}
    \Text(22,0)[]{\normalsize{\Black{$\Bigg\{$}}}
    \DashLine(33,3)(33,-10){2}\DashLine(33,3)(35,3){2}
    \DashLine(33,-10)(40,-10){2}
    \Cross(38,17)\Cross(38,3)
    \DashCArc(38,10)(7,270,90){3}
    \ArrowArcn(45,0)(10,90,0)
    \DashCArc(45,0)(10,270,0){3}
    \Cross(45,-10)
    \ArrowLine(55,0)(70,0)
  \Vertex(70,0){1.5}
  \Text(58,4)[]{\normalsize{\Black{$y$}}}
  \Text(48,14)[]{\normalsize{\Black{$z$}}}
  \Text(71,5)[]{\normalsize{\Black{$x_2$}}}
    \Text(80,0)[]{\normalsize{\Black{$\Bigg\}_{\!2}$}}}
\end{pic}
\right\rangle =
  \begin{pic}{62}{30}{(-5,-15)}
    \SetWidth{0.5}
    \SetColor{Black}
    \Vertex(0,0){1.5}
    \DashLine(0,0)(18,0){3}\Cross(9,0)
    \ArrowArcn(28,0)(10,180,0)
    \DashCArc(28,0)(10,180,0){3}\Cross(28,-10)
    \ArrowLine(38,0)(56,0)
    \Vertex(56,0){1.5}
    \Text(2,5)[]{\normalsize{\Black{$x_1$}}}
    \Text(15,-4)[]{\normalsize{\Black{$z$}}}
    \Text(41,-4)[]{\normalsize{\Black{$y$}}}
    \Text(57,5)[]{\normalsize{\Black{$x_2$}}}
  \end{pic}
\end{equation}
(where again twisting the whole diagram upside-down would return an
equivalent one).

In all the previous graphs there are two vertices and two ways of
placing the dashed lines. This yields then a global symmetry factor 
of $\frac{1}{2}$.
The two-point correlation function therefore reads 
\begin{align}
  &\mean{\phi(x_1)\phi(x_2)} = \mean{\phi_0(x_1)\phi_0(x_2)} 
  + \frac{\lambda^2}{2} \!\!\int \dd y a_y^3 \,\dd z a_z^3 \notag \\
  &\quad\times\Big[G_R(x_1,y)G_R(y,z)
  \mean{\{\phi_0(y),\phi_0(z)\}} \!\mean{\phi_0(z)\phi_0(x_2)} \notag \\ 
  &\quad+ G_R(x_2,y)G_R(y,z)
  \mean{\phi_0(x_1)\phi_0(z)}\!\mean{\{\phi_0(y),\phi_0(z)\}} \notag \\
  &\quad+ G_R(x_1,y)G_R(x_2,z) \!\mean{\phi_0(y)\phi_0(z)}^2\Big] 
\end{align}
up to terms of order $\mathcal{O}(\lambda^4)$.

From the previous analysis we might be tempted to say that we can 
associate to every internal dashed line the expectation value of 
an anticommutator.
Unfortunately this is not always the case, but more complicated
symmetrization procedures can occur.
As an example, consider the fourth order graph
\begin{equation}
\label{gr2p4th}
  \begin{pic}{60}{40}{(-5,-20)}
    \SetWidth{0.5}
    \SetColor{Black}
    \Vertex(-3,0){1.5}
    \ArrowLine(15,0)(-3,0)
    \ArrowArc(27.5,0)(12.5,90,180)
    \ArrowArc(27.5,0)(12.5,0,90)
    \ArrowArcn(27.5,0)(12.5,270,180)
    \DashCArc(27.5,0)(12.5,270,0){3}
    \Text(36,-9)[]{\textbf{\scriptsize{+}}}
    \DashLine(27.5,-12.5)(27.5,12.5){3}\Cross(27.5,0)
    \DashLine(40,0)(58,0){3}\Cross(49,0)
    \Vertex(58,0){1.5}
    \Text(-1,5)[]{\normalsize{\Black{$x_1$}}}
    \Text(59,5)[]{\normalsize{\Black{$x_2$}}}
    \Text(12,-5)[]{\normalsize{\Black{$y$}}}
    \Text(27.5,16)[]{\normalsize{\Black{$w$}}}
    \Text(43,-5)[]{\normalsize{\Black{$z$}}}
    \Text(27,-17)[]{\normalsize{\Black{$u$}}}
  \end{pic}
\end{equation}
contributing to the two-point correlation function. This diagram
comes from the expectation value
\begin{equation}
\left\langle
  \begin{pic}{105}{50}{(-10,-25)}
    \SetWidth{0.5}
    \SetColor{Black}
    \Text(-7,0)[]{\normalsize{\Black{$\Bigg\{$}}}
    \Vertex(0,0){1.2}
    \ArrowLine(20,0)(0,0)
    \ArrowArc(32,0)(12,90,180)
    \ArrowArc(40,12)(8,90,180)\ArrowArcn(32,0)(12,270,180)
    \DashCArc(40,12)(8,180,270){3}
    \DashCArc(40,-12)(8,90,270){3}
    \Cross(40,-20)\Cross(40,-4)\Cross(40,4)
    \DashLine(43,4)(45,4){2}\DashLine(45,4)(45,-4){2}
    \DashLine(43,-4)(45,-4){2}
    \DashCArc(45,20)(5,90,270){2}
    \Cross(45,25)\Cross(45,15)
    \DashLine(48,15)(50,15){2}\DashLine(44,-20)(50,-20){2}
    \DashLine(50,15)(50,-20){2}
    \Text(2,5)[]{\normalsize{\Black{$x_1$}}}
    \Text(17,5)[]{\normalsize{\Black{$y$}}}
    \Text(28,16)[]{\normalsize{\Black{$w$}}}
    \Text(37,24)[]{\normalsize{\Black{$z$}}}
    \Text(29,-16)[]{\normalsize{\Black{$u$}}}
    \DashLine(48,25)(54,25){2}\DashLine(54,25)(54,0){2}
    \DashLine(54,0)(65,0){2}
    \Text(64,0)[]{\normalsize{\Black{$\Bigg\}_{\!2,2}$}}}
    \Cross(72,0)\DashLine(72,0)(88,0){3}
    \Vertex(88,0){1.3}
    \Text(90,5)[]{\normalsize{\Black{$x_2$}}}
  \end{pic} 
\right\rangle,
\end{equation}
that is of a subset of $\df_4(x_1)$ containing four terms and of 
$\phi_0(x_2)$. The total free field content in the expectation value is
\begin{gather}
  \mean{\Big\{\phi_0^2(u),\{\phi_0(w),\phi_0^2(z)\}\!\Big\}\phi_0(x_2)},
  \\[-5pt]
\begin{pic}{110}{0}{(-5,0)}
  \Line(3,3)(3,5)\DashLine(3,3)(35,3){3}\Line(35,3)(35,5)
  \Line(0,0)(0,5)\DashLine(0,0)(60,0){3}\Line(60,0)(60,5)
  \Line(63,0)(63,5)\DashLine(63,0)(95,0){3}\Line(95,0)(95,5)
\end{pic}
\notag
\end{gather}
where the internal anticommutator is associated with the vertex $w$,
the external one with the vertex $y$, and the dashed lines indicate 
the pairings that we are actually considering in \eqref{gr2p4th}.

The peculiarity here is the fact that the graph \eqref{gr2p4th} does 
not pair $\phi_0(w)$ with $\phi_0(z)$ (such pair is contained in a 
different graph). The only pairings are $\phi_0(u)$ with 
$\phi_0(w)$ or with $\phi_0(z)$, and are not affected by 
the order of the latter fields. Therefore, the internal anticommutator 
does not contain any dashed line and merely introduces a factor of 2. 
The only anticommutator affecting the dashed lines is the external one, 
that reverses both lines simultaneously: indeed, the result is 
$2[\mean{\phi_w\phi_u}\!\mean{\phi_z\phi_u} +
\mean{\phi_u\phi_w}\!\mean{\phi_u\phi_z}]$. This differs from our
naive expectation of the product of two anticommutators by the presence 
of a factor of 2 and the absence of mixed terms.
We also have a factor of $(1/2)^4$ from the vertices and a combinatorial
factor of 4 counting the equivalent dispositions of the dashed lines.
The net contribution of \eqref{gr2p4th} then is
\begin{align}
  \lambda^4 \!\!
  &\int\dd y\,a_y^3 \dd w\,a_w^3 \dd z\,a_z^3 \dd u\,a_u^3 \,
  G_R(x_1,y) G_R(y,w) \notag \\
  &\times G_R(w,z) G_R(y,u)  \frac{1}{2}\Big[
  \mean{\phi_0(w)\phi_0(u)}\!\mean{\phi_0(z)\phi_0(u)} \notag \\
  &+ \mean{\phi_0(u)\phi_0(w)}\!\mean{\phi_0(u)\phi_0(z)} \Big]\!
  \mean{\phi_0(z)\phi_0(x_2)},
\end{align}
and as said above it contains only one symmetrization involving two
internal dashed lines at the same time.

In general, all intermediate vertices of the trees (\ie vertices with 
at least one outgoing solid line besides the ingoing one) are associated 
to anticommutators that exchanges the $n-1$ slots of the outgoing 
lines\footnote{In open perturbative diagrams not all intermediate 
vertices are symmetrized: vertices with a symmetric sub-tree (\ie 
indistinguishable outgoing solid lines) do not bear any anticommutator. 
However, in closed graphs one must sum all the possible ways to connect 
external dashed lines to them, and this provide an effective 
symmetrization of such vertices (see App.~\ref{app1} for details)}. 
Any dashed line connected with both ends to two of the $n-1$ slots 
of a given vertex $z$ will thus be symmetrized by the anticommutator.
When acting on a solid line, the anticommutator acts on the whole 
sub-tree starting from that line: therefore not only the dashed lines 
directly attached to the vertex $z$, but also all those connecting 
vertices in the sub-tree of two outgoing solid lines, will be symmetrized
at the same time.
On the other hand, dashed lines having the second end attached to a 
lower vertex $y$ (or to any other of its outgoing solid lines) are 
reversed by the anticommutator in $y$ and not by the one in $z$.

We can now classify the behavior of a generic dashed line on
the basis of its position in the graph.
All closed graphs are constructed with open trees of solid lines 
starting from each external point, connected with dashed lines that 
close the diagram. 
As we have seen, the latter can be internal or external. Internal 
dashed lines connect vertices in the same tree, and are therefore 
necessarily part of a loop with only one dashed line (with all the 
solid lines of the loop belonging to same perturbative tree). We will 
refer to this unique loop as to the symmetrization loop of the line.
Dashed lines for which such a loop does not exist are of the external 
type, and always connect vertices in different perturbative trees.
The extreme case of external dashed line is when the line is directly
attached to an external point.
Any dashed line connecting two points $x$ and $y$ is related to the 
two-field expectation value of $\phi_0(x)$ and $\phi_0(y)$, but the
order of the fields depends on which kind of line we are considering.

From what said above, it follows that internal dashed lines are
symmetrized by the anticommutator associated to the lowest vertex of 
their symmetrization loop (\ie the one closest to the external point).
Whenever two or more dashed lines are symmetrized by the same vertex, 
the end points of all the lines must be exchanged simultaneously. 
For example, in the case of \eqref{gr2p4th}, the two loops are 
$y$-$w$-$u$-$y$ and $y$-$w$-$z$-$u$-$y$ and in both cases the 
lowest vertex is $y$. In general, when $k$ lines are symmetrized 
by the same vertex one needs to introduce the factor (for a $\phi^3$
potential)
\begin{equation}
\label{simm}
  \frac{1}{2}\left[\prod_{i=1}^l\mean{\phi_0(y_i)\phi_0(z_i)} +
  \prod_{i=1}^k\mean{\phi_0(z_i)\phi_0(y_i)}\right],
\end{equation}
where for convenience we introduced the normalization $\frac{1}{2}$
which will be useful in the following.

In the case of a generic $\phi^n$ interaction one has $(n-1)!$ 
permutations of the $n-1$ outgoing lines. Any vertex $y_i$ attached 
to the line $l_{y_i}$ will be associated by each permutation 
$p\in S_{n-1}$ to the line $p(l_{y_i})$. The order of two fields
$\phi_0(y_i)$ and $\phi_0(z_i)$ is then reversed when the permutation
changes the sign of $p(l_{y_i})-p(l_{z_i})$.
The generalization to the previous formula to a generic interaction
is therefore immediate, and reads
\begin{align}
\label{simmgen}
  \frac{1}{(n-1)!}&\sum_p\prod_{i=1}^k
  \bigg\{\vartheta\Big[p\left(l_{y_i}\right)-p\left(l_{z_i}\right)\!\Big]\!
  \mean{\phi_0(y_i)\phi_0(z_i)} \notag \\
  &+ \vartheta\Big[p\left(l_{z_i}\right)-p\left(l_{y_i}\right)\!\Big]\!
  \mean{\phi_0(z_i)\phi_0(y_i)}\!\bigg\},
\end{align}
where again we introduced a normalization factor by hand.

Dashed lines symmetrized by different vertices enter as a separate 
factor, so that there will be as many factors of $1/(n-1)!$ as 
symmetrizing vertices.
On the other hand, external dashed lines do not need to be symmetrized:
the order of the free fields in the expectation value is fixed, and
is the same as the order the two external points to which they are 
connected have in the complete correlation function.


\begin{figure*}
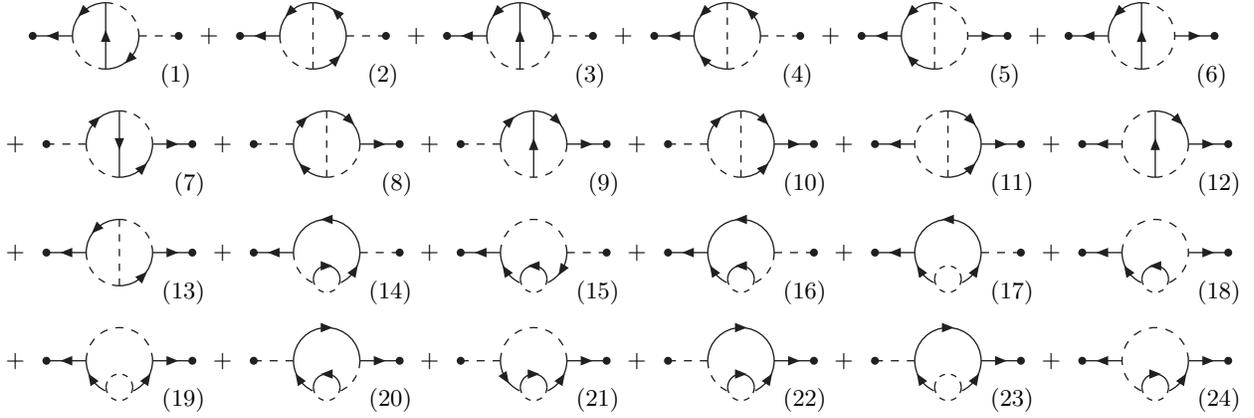

\begin{minipage}{\textwidth}
  \begin{pic}{65}{40}{(-5,-20)} 
    \SetWidth{0.5}
    \SetColor{Black}
    \Vertex(0,0){1.5}
    \ArrowLine(15,0)(0,0)
    \ArrowArc(27.5,0)(12.5,90,180)\ArrowArcn(27.5,0)(12.5,0,270)
    \DashCArc(27.5,0)(12.5,0,90){3}\DashCArc(27.5,0)(12.5,180,270){3}
    \ArrowLine(27.5,-12.5)(27.5,12.5)
    \DashLine(40,0)(55,0){3}
    \Vertex(55,0){1.5}
  \end{pic}\raisebox{-15pt}{\makebox[0pt][r]{(1)}} + 
  \begin{pic}{65}{40}{(-5,-20)}
    \SetWidth{0.5}
    \SetColor{Black}
    \Vertex(0,0){1.5}
    \ArrowLine(15,0)(0,0)
    \ArrowArc(27.5,0)(12.5,270,0)
    \ArrowArc(27.5,0)(12.5,0,90)\ArrowArc(27.5,0)(12.5,90,180)
    \DashCArc(27.5,0)(12.5,180,270){3}
    \DashLine(27.5,-12.5)(27.5,12.5){3}
    \DashLine(40,0)(55,0){3}
    \Vertex(55,0){1.5}
  \end{pic}\raisebox{-15pt}{\makebox[0pt][r]{(2)}} +
  \begin{pic}{65}{40}{(-5,-20)}
    \SetWidth{0.5}
    \SetColor{Black}
    \Vertex(0,0){1.5}
    \ArrowLine(15,0)(0,0)
    \ArrowArc(27.5,0)(12.5,0,90)\ArrowArc(27.5,0)(12.5,90,180) 
    \DashCArc(27.5,0)(12.5,180,0){3}
    \ArrowLine(27.5,-12.5)(27.5,12.5)
    \DashLine(40,0)(55,0){3}
    \Vertex(55,0){1.5}
  \end{pic}\raisebox{-15pt}{\makebox[0pt][r]{(3)}} +
  \begin{pic}{65}{40}{(-5,-20)}
    \SetWidth{0.5}
    \SetColor{Black}
    \Vertex(0,0){1.5}
    \ArrowLine(15,0)(0,0)
    \ArrowArc(27.5,0)(12.5,0,90)\ArrowArc(27.5,0)(12.5,90,180)
    \ArrowArcn(27.5,0)(12.5,270,180)
    \DashCArc(27.5,0)(12.5,270,0){3}
    \DashLine(27.5,-12.5)(27.5,12.5){3}
    \DashLine(40,0)(55,0){3}
    \Vertex(55,0){1.5}
  \end{pic}\raisebox{-15pt}{\makebox[0pt][r]{(4)}} +
  \begin{pic}{65}{40}{(-5,-20)}
    \SetWidth{0.5}
    \SetColor{Black}
    \Vertex(0,0){1.5}
    \ArrowLine(15,0)(0,0)
    \ArrowArc(27.5,0)(12.5,90,180)
    \ArrowArcn(27.5,0)(12.5,270,180)
    \DashCArc(27.5,0)(12.5,270,90){3}
    \DashLine(27.5,-12.5)(27.5,12.5){3}
    \ArrowLine(40,0)(55,0)
    \Vertex(55,0){1.5}
  \end{pic}\raisebox{-15pt}{\makebox[0pt][r]{(5)}} +
  \begin{pic}{65}{40}{(-5,-20)}
    \SetWidth{0.5}
    \SetColor{Black}
    \Vertex(0,0){1.5}
    \ArrowLine(15,0)(0,0)
    \ArrowArc(27.5,0)(12.5,90,180) \DashCArc(27.5,0)(12.5,180,270){3}
    \DashCArc(27.5,0)(12.5,270,90){3}
    \ArrowLine(27.5,-12.5)(27.5,12.5)
    \ArrowLine(40,0)(55,0)
    \Vertex(55,0){1.5}
  \end{pic}\raisebox{-15pt}{\makebox[0pt][r]{(6)}} \\ +
  \begin{pic}{65}{40}{(-5,-20)}
    \SetWidth{0.5}
    \SetColor{Black}
    \Vertex(0,0){1.5}
    \DashLine(0,0)(15,0){3}
    \ArrowArcn(27.5,0)(12.5,180,90)\ArrowArc(27.5,0)(12.5,270,0)
    \DashCArc(27.5,0)(12.5,0,90){3}\DashCArc(27.5,0)(12.5,180,270){3}
    \ArrowLine(27.5,12.5)(27.5,-12.5)
    \ArrowLine(40,0)(55,0)
    \Vertex(55,0){1.5}
  \end{pic}\raisebox{-15pt}{\makebox[0pt][r]{(7)}} + 
  \begin{pic}{65}{40}{(-5,-20)}
    \SetWidth{0.5}
    \SetColor{Black}
    \Vertex(0,0){1.5}
    \DashLine(0,0)(15,0){3}
    \ArrowArcn(27.5,0)(12.5,270,180)
    \ArrowArcn(27.5,0)(12.5,180,90)\ArrowArcn(27.5,0)(12.5,90,0)
    \DashCArc(27.5,0)(12.5,270,0){3}
    \DashLine(27.5,-12.5)(27.5,12.5){3}
    \ArrowLine(40,0)(55,0)
    \Vertex(55,0){1.5}
  \end{pic}\raisebox{-15pt}{\makebox[0pt][r]{(8)}} +
  \begin{pic}{65}{40}{(-5,-20)}
    \SetWidth{0.5}
    \SetColor{Black}
    \Vertex(0,0){1.5}
    \DashLine(0,0)(15,0){3}
    \ArrowArcn(27.5,0)(12.5,180,90)\ArrowArcn(27.5,0)(12.5,90,0)
    \DashCArc(27.5,0)(12.5,180,0){3}
    \ArrowLine(27.5,-12.5)(27.5,12.5)
    \ArrowLine(40,0)(55,0)
    \Vertex(55,0){1.5}
  \end{pic}\raisebox{-15pt}{\makebox[0pt][r]{(9)}} +
  \begin{pic}{65}{40}{(-5,-20)}
    \SetWidth{0.5}
    \SetColor{Black}
    \Vertex(0,0){1.5}
    \DashLine(0,0)(15,0){3}
    \ArrowArc(27.5,0)(12.5,270,0)
    \ArrowArcn(27.5,0)(12.5,180,90)\ArrowArcn(27.5,0)(12.5,90,0)
    \DashCArc(27.5,0)(12.5,180,270){3}
    \DashLine(27.5,-12.5)(27.5,12.5){3}
    \ArrowLine(40,0)(55,0)
    \Vertex(55,0){1.5}
  \end{pic}\raisebox{-15pt}{\makebox[0pt][r]{(10)}} +
  \begin{pic}{65}{40}{(-5,-20)}
    \SetWidth{0.5}
    \SetColor{Black}
    \Vertex(0,0){1.5}
    \ArrowLine(15,0)(0,0)
    \DashCArc(27.5,0)(12.5,90,270){3}
    \ArrowArc(27.5,0)(12.5,270,0)\ArrowArcn(27.5,0)(12.5,90,0)
    \DashLine(27.5,-12.5)(27.5,12.5){3}
    \ArrowLine(40,0)(55,0)
    \Vertex(55,0){1.5}
  \end{pic}\raisebox{-15pt}{\makebox[0pt][r]{(11)}} +
  \begin{pic}{65}{40}{(-5,-20)}
    \SetWidth{0.5}
    \SetColor{Black}
    \Vertex(0,0){1.5}
    \ArrowLine(15,0)(0,0)
    \DashCArc(27.5,0)(12.5,90,270){3}
    \ArrowArcn(27.5,0)(12.5,90,0)
    \DashCArc(27.5,0)(12.5,270,0){3}
    \ArrowLine(27.5,-12.5)(27.5,12.5)
    \ArrowLine(40,0)(55,0)
    \Vertex(55,0){1.5}
  \end{pic}\raisebox{-15pt}{\makebox[0pt][r]{(12)}} \\ +
  \begin{pic}{65}{40}{(-5,-20)}
    \SetWidth{0.5}
    \SetColor{Black}
    \Vertex(0,0){1.5}
    \ArrowLine(15,0)(0,0)
    \ArrowArc(27.5,0)(12.5,90,180)
    \DashCArc(27.5,0)(12.5,180,270){3}
    \ArrowArc(27.5,0)(12.5,270,0) \DashCArc(27.5,0)(12.5,0,90){3}
    \DashLine(27.5,-12.5)(27.5,12.5){3}
    \ArrowLine(40,0)(55,0)
    \Vertex(55,0){1.5}
  \end{pic}\raisebox{-15pt}{\makebox[0pt][r]{(13)}} +
  \begin{pic}{65}{40}{(-5,-20)}
    \SetWidth{0.5}
    \SetColor{Black}
    \Vertex(0,0){1.5}
    \ArrowLine(15,0)(0,0)
    \ArrowArc(27.5,0)(12.5,295,0)
    \ArrowArc(27.5,0)(12.5,0,180)
    \DashCArc(27.5,0)(12.5,180,245){3}
    \ArrowArcn(27.5,-10)(5,180,0) \DashCArc(27.5,-10)(5,180,0){2}
    \DashLine(40,0)(55,0){3}
    \Vertex(55,0){1.5}
  \end{pic}\raisebox{-15pt}{\makebox[0pt][r]{(14)}} +
  \begin{pic}{65}{40}{(-5,-20)}
    \SetWidth{0.5}
    \SetColor{Black}
    \Vertex(0,0){1.5}
    \ArrowLine(15,0)(0,0)
    \DashCArc(27.5,0)(12.5,0,180){3}
    \ArrowArcn(27.5,0)(12.5,245,180)\ArrowArcn(27.5,0)(12.5,0,295)
    \ArrowArc(27.5,-10)(5,0,180) \DashCArc(27.5,-10)(5,180,0){2}
    \DashLine(40,0)(55,0){3}
    \Vertex(55,0){1.5}
  \end{pic}\raisebox{-15pt}{\makebox[0pt][r]{(15)}} +
  \begin{pic}{65}{40}{(-5,-20)}
    \SetWidth{0.5}
    \SetColor{Black}
    \Vertex(0,0){1.5}
    \ArrowLine(15,0)(0,0)
    \ArrowArc(27.5,0)(12.5,0,180)\ArrowArcn(27.5,0)(12.5,245,180)
    \DashCArc(27.5,0)(12.5,295,0){3}
    \ArrowArc(27.5,-10)(5,0,180) \DashCArc(27.5,-10)(5,180,0){2}
    \DashLine(40,0)(55,0){3}
    \Vertex(55,0){1.5}
  \end{pic}\raisebox{-15pt}{\makebox[0pt][r]{(16)}} +
  \begin{pic}{65}{40}{(-5,-20)}
    \SetWidth{0.5}
    \SetColor{Black}
    \Vertex(0,0){1.5}
    \ArrowLine(15,0)(0,0)
    \ArrowArc(27.5,0)(12.5,0,180)\ArrowArcn(27.5,0)(12.5,245,180)
    \ArrowArc(27.5,0)(12.5,295,0)
    \DashCArc(27.5,-10)(5,0,180){2} \DashCArc(27.5,-10)(5,180,0){2}
    \DashLine(40,0)(55,0){3}
    \Vertex(55,0){1.5}
  \end{pic}\raisebox{-15pt}{\makebox[0pt][r]{(17)}} +
  \begin{pic}{65}{40}{(-5,-20)}
    \SetWidth{0.5}
    \SetColor{Black}
    \Vertex(0,0){1.5}
    \ArrowLine(15,0)(0,0)
    \DashCArc(27.5,0)(12.5,295,180){3} \ArrowArcn(27.5,0)(12.5,245,180)
    \ArrowArc(27.5,-10)(5,0,180) \DashCArc(27.5,-10)(5,180,0){2}
    \ArrowLine(40,0)(55,0)
    \Vertex(55,0){1.5}
  \end{pic}\raisebox{-15pt}{\makebox[0pt][r]{(18)}} \\ +
  \begin{pic}{65}{40}{(-5,-20)}
    \SetWidth{0.5}
    \SetColor{Black}
    \Vertex(0,0){1.5}
    \ArrowLine(15,0)(0,0)
    \DashCArc(27.5,0)(12.5,0,180){3}
    \ArrowArcn(27.5,0)(12.5,245,180) \ArrowArc(27.5,0)(12.5,295,0)
    \DashCArc(27.5,-10)(5,0,180){2} \DashCArc(27.5,-10)(5,180,0){2}
    \ArrowLine(40,0)(55,0)
    \Vertex(55,0){1.5}
  \end{pic}\raisebox{-15pt}{\makebox[0pt][r]{(19)}} +
  \begin{pic}{65}{40}{(-5,-20)}
    \SetWidth{0.5}
    \SetColor{Black}
    \Vertex(0,0){1.5}
    \DashLine(0,0)(15,0){3}
    \ArrowArcn(27.5,0)(12.5,245,180)
    \ArrowArcn(27.5,0)(12.5,180,0)\DashCArc(27.5,0)(12.5,295,0){3}
    \ArrowArc(27.5,-10)(5,0,180) \DashCArc(27.5,-10)(5,180,0){2}
    \ArrowLine(40,0)(55,0)
    \Vertex(55,0){1.5}
  \end{pic}\raisebox{-15pt}{\makebox[0pt][r]{(20)}} +
  \begin{pic}{65}{40}{(-5,-20)}
    \SetWidth{0.5}
    \SetColor{Black}
    \Vertex(0,0){1.5}
    \DashLine(0,0)(15,0){3}
    \DashCArc(27.5,0)(12.5,0,180){3}
    \ArrowArc(27.5,0)(12.5,180,245)\ArrowArc(27.5,0)(12.5,295,0)
    \ArrowArcn(27.5,-10)(5,180,0) \DashCArc(27.5,-10)(5,180,0){2}
    \ArrowLine(40,0)(55,0)
    \Vertex(55,0){1.5}
  \end{pic}\raisebox{-15pt}{\makebox[0pt][r]{(21)}} +
  \begin{pic}{65}{40}{(-5,-20)}
    \SetWidth{0.5}
    \SetColor{Black}
    \Vertex(0,0){1.5}
    \DashLine(0,0)(15,0){3}
    \ArrowArc(27.5,0)(12.5,295,0)
    \ArrowArcn(27.5,0)(12.5,180,0)
    \DashCArc(27.5,0)(12.5,180,245){3}
    \ArrowArcn(27.5,-10)(5,180,0) \DashCArc(27.5,-10)(5,180,0){2}
    \ArrowLine(40,0)(55,0)
    \Vertex(55,0){1.5}
  \end{pic}\raisebox{-15pt}{\makebox[0pt][r]{(22)}} +
  \begin{pic}{65}{40}{(-5,-20)}
    \SetWidth{0.5}
    \SetColor{Black}
    \Vertex(0,0){1.5}
    \DashLine(0,0)(15,0){3}
    \ArrowArc(27.5,0)(12.5,295,0)
    \ArrowArcn(27.5,0)(12.5,180,0)
    \ArrowArcn(27.5,0)(12.5,245,180)
    \DashCArc(27.5,-10)(5,0,180){2} \DashCArc(27.5,-10)(5,180,0){2}
    \ArrowLine(40,0)(55,0)
    \Vertex(55,0){1.5}
  \end{pic}\raisebox{-15pt}{\makebox[0pt][r]{(23)}} +
  \begin{pic}{65}{40}{(-5,-20)}
    \SetWidth{0.5}
    \SetColor{Black}
    \Vertex(0,0){1.5}
    \ArrowLine(15,0)(0,0)
    \DashCArc(27.5,0)(12.5,0,245){3} \ArrowArc(27.5,0)(12.5,295,0)
    \ArrowArcn(27.5,-10)(5,180,0) \DashCArc(27.5,-10)(5,180,0){2}
    \ArrowLine(40,0)(55,0)
    \Vertex(55,0){1.5}
  \end{pic}\raisebox{-15pt}{\makebox[0pt][r]{(24)}} 
\end{minipage}
\caption{\label{fig1} Fourth order contributions to the two-point 
correlation function (for ease of notation, we have removed crosses from dashed lines, without changing their mathematical meaning). The symmetry factor is 1 for all graphs but (5), 
(11), (17), (18), (23) and (24), for which it is $\frac{1}{2}$. In graphs 
(4) and (10) the internal dashed lines share the same symmetrizing vertex
and are symmetrized together, yielding 
$\frac{1}{2}\left[\mean{\phi_0(u)\phi_0(w)}\mean{\phi_0(u)\phi_0(z)} 
+ \mean{\phi_0(w)\phi_0(u)}\mean{\phi_0(z)\phi_0(u)}\right]$, as well 
as in (17) and (23) where they give $\frac{1}{2}\left[
\mean{\phi_0(z)\phi_0(w)}^2 + \mean{\phi_0(w)\phi_0(z)}^2\right]$.
In all the other graphs all internal dashed lines are symmetrized
individually, accounting $\frac{1}{2}\left[\mean{\phi_0(y_i)\phi_0(y_j)}
+\mean{\phi_0(y_j)\phi_0(y_i)}\right]$ for each line.} 
\end{figure*}

\section{Symmetry factors}
\label{Symmetry}

We now want to understand how to compute the overall symmetry factor 
for each closed graph, without going through the complicated procedure 
of connecting all the equivalent open diagrams.

We notice that we assigned in \eqref{simm} the same normalization 
$\frac{1}{2}$ to all groups of internal dashed lines symmetrized by 
a single vertex.
As explained in App.~\ref{app1}, this factor needs to be introduced 
when the dashed line connects the ends of two symmetric solid lines
stemming from a vertex. In this case the vertex does not have a
proper anticommutator and the symmetrization of the line(s) is provided
by the different ways of connecting other lines to its end points.
The combinatorial factor of the additional lines doubles
after the symmetrization (the two end points are now equivalent), 
and the $\frac{1}{2}$ is needed to reproduce the correct result.
When instead the symmetrization is produced by a real anticommutator
(like in all the examples throughout the current Section), 
the $\frac{1}{2}$ factor does not arise and must be introduced by hand:
the associated vertex must then be given a factor of 2 to compensate.
Finally, vertices with ``idle'' anticommutators also introduce a
factor of 2.

It follows that any vertex with just one outgoing solid line carries
a factor of 2, while vertices with two do not carry any additional
factor since they are symmetric (they could still have of course
a combinatorial factor if the outgoing lines can be equivalently
attached to several other vertices, but no factor for the exchange
of the lines). Final vertices (with only outgoing dashed lines)
carry the standard combinatorial factor counting the equivalent 
dispositions of the dashed lines.
All these factors are computed holding the solid lines fixed, and must 
multiply the factors of $\frac{1}{2}$ introduced by the vertices, which 
amount to an overall factor of $2^{-N}$ for all $N$-th order graphs.

This procedure is equivalent to the one used in standard QFT, where
a factor of $(1/N!)(1/3!)^N$ is assigned to any graph, and afterward 
multiplied by the number of equivalent configurations. Indeed, the 
number of possibilities just for placing the $N$ solid lines (that do
not form any loop) automatically 
contains a factor of $N!3^N$, which gives back the original $2^{-N}$.
The only difference from the standard computation of symmetry factors
is that one has to distinguish between the two types of lines:
when placing a solid line in a vertex containing also a dashed line
(and vice versa) one must count the legs of both the starting and 
arriving vertex.
This introduces extra combinatorial factors, which exactly account
for the factors of two we need in asymmetric vertices.

Let us try to clarify the outlined procedure with a couple of 
examples. In the graph in Eq.~\eqref{gr2p4th} [the same as graph 
(4) of Fig.~\ref{fig1}], which starts from $(1/4!)(1/3!)^4$, we have 
$4\cdot 3$ ways to place the $x_1y$ line, $3\cdot 3$ for $yw$ (we do 
not count the starting legs in $y$ because $yw$ and $yu$ are symmetric)
$2\cdot3$ for $yu$, 3 for $x_2z$, $2\cdot4$ for $zw$ (2 starting
point in $z$ since $zw$ is different from $zu$, and 4 equivalent
arriving points, 2 in $z$ and 2 in $u$), 2 for $zu$ and 1 for $uw$.
The total symmetry factor is then 1, and the two internal dashed lines
are symmetrized together introducing a single normalization factor 
of $\frac{1}{2}$. 
The graph in \eqref{gr2p_2} has a symmetry factor of 1: there are 
$3\cdot2$ possibilities for $x_1y$, 3 for $x_2z$, $2\cdot2$ for the 
solid $yz$ (the two lines in the loop are different and thus one must 
count also the starting legs), while the dashed $yz$ is fixed. These
exactly compensate the $(1/2)(1/3!)^2$. In addition, there is one 
normalization factor of $\frac{1}{2}$.
The same calculation holds for the graph \eqref{gr2p_3}, while the one 
in \eqref{gr2p} has a symmetry factor of $\frac{1}{2}$ (the two lines 
in the loop are indistinguishable) and no symmetrization. 
Some more examples are given in Fig.~\ref{fig1}.

In general, when the symmetrization of the internal dashed lines
is normalized as in \eqref{simmgen} the total symmetry factor $S$ of
each $N$-th order graph is given by
\begin{equation}
\label{factor}
  S = \frac{P}{N!(n!)^N},
\end{equation}
where $P$ is the number of equivalent ways to place the lines.
This is the same procedure to compute the symmetry factors in standard
QFT. However, unlike in the standard case, when counting the equivalent
configurations one must distinguish solid and dashed lines.
The resulting symmetry factor is therefore generically larger than
the usual one, though still smaller than unity [and does not include
all the factors of $1/(n-1)!$ from the normalization of symmetrized
dashed lines].


\section{Summary and Conclusions}
\label{conclusion}

We are now able to summarize all the results obtained in the previous
sections.
An iterative solution of the equation of motion for the interacting 
operator field was given, where higher order corrections are expressed in 
terms of lower order ones and retarded Green functions. 
Correlation functions obtained through this procedure were shown to be exactly equivalent to the local $n$-point correlations computed in the \emph{in-in} formalism along a close time path, even though this concept was never explictely needed. In our case, locality is an immediate consequence of the use of only retarded Green functions.
The iterative procedure also suggested a Feynman-like picture of the 
various contributions, with a modified
interpretation of the lines of the diagrams.

Since two different 
kinds of line - solid or dashed - can be used, inequivalent versions of 
the same diagram can be generated simply by rearranging the disposition 
of the lines. Therefore, as a first step one has to draw all such 
possible versions, finding out all the allowed configurations of solid 
and dashed lines. Secondly, the correct mathematical expression must be 
associated to each element of the graph\footnote{The diagrammatic 
expansion presented here is somewhat similar to the one described in 
Ref.~\cite{Chou}, which however also involves advanced Green functions
and thus it is clearly not the same. It turns out that the formulation
given in this paper is slightly more efficient, as it requires less
diagrams to describe the same correlation functions. 
Using only retarded Green functions also makes each diagram explicitely 
local, without relying on cancellations between different diagrams
to preserve the locality of the final result}.

The prescriptions to build an allowed configuration are the following:
\begin{itemize}
\item all solid lines form connected trees starting from an external 
point (there cannot be isolated solid lines)
\item all vertices are reached by at least one solid line
\item the total number of solid lines equals the number of 
vertices
\item there are no solid loops, nor connected trees reaching different
external points
\item all the remaining lines are dashed
\end{itemize}
The sum of all the allowed permutations of solid and dashed lines 
preserving the shape of the graph represents the total contribution of 
that Feynman graph in the \emph{in-in} formalism (see Fig.\ref{fig1}).

In turn, dashed lines belong to two different types: internal (between 
vertices in the same tree) or external (connecting different trees).
By definition, for any internal dashed line there exists a unique 
loop having that as the only dashed line, which we have called the 
symmetrization loop. The lowest vertex of the loop, \ie the one being
the closest to the root of the tree, is the symmetrization vertex.
In order to reconstruct the mathematical content of each graph, 
the following rules apply:
\begin{itemize}
\item each vertex brings an integration factor 
$-\lambda\!\int\dd y\, a^3$
\item each solid line corresponds to a retarded Green function
$G_{\!R}(y,z)$
\item an external dashed line represents the expectation value 
$\mean{\phi_0(y)\phi_0(z)}$ or $\mean{\phi_0(z)\phi_0(y)}$. The order 
of the fields is fixed by the order of the respective trees in the 
correlation function
\item an internal dashed line not sharing the symmetrization vertex
contains the symmetrized expectation value $\frac{1}{2}\left[
\mean{\phi_0(y)\phi_0(z)} + \mean{\phi_0(z)\phi_0(y)}\right]$
\item several internal dashed lines sharing the same symmetrization 
vertex are symmetrized simultaneously using Eq.~\eqref{simm} or
\eqref{simmgen}
\item all graphs are multiplied by the symmetry factor \eqref{factor},
computed distinguishing between solid and dashed lines
\end{itemize}

The above prescriptions introduce a useful practical method to reconnect 
the reformulation of the \emph{in-in} formalism given in \cite{Weinberg}
with the Feynman diagrams, providing a quick and efficient tool to 
reconstruct all the perturbative terms without any calculation.
The modified Feynman rules describe loop correlation functions in terms
of retarded Green functions and two-field expectation values. This
immediately reflects the decomposition outlined in \cite{Weinberg} of
loop contributions into free fields and commutators of free fields.
It thus allows to study in an easy way the late time behavior of these
contributions, and analyze their dependence on powers of $\log a(t)$.
A precise calculation is needed to investigate the possibility of a 
resummation of all the powers of $\log a(t)$, which might eventually 
produce a dependence on some finite power of the scale factor.

This modified diagrammatic formalism can also be used to study the 
divergences appearing in the loops. Although the integrals over time 
are basically understood, most diagrams are in fact plagued by
divergences in the integrals over the three-momentum \cite{Woodard}.
A correct understanding of these divergences is therefore necessary to 
investigate the behavior of the loop contributions, and to estimate 
their relevance with respect to the tree-level terms. 
The ultimate goal to achieve would be the removal of the divergences
and the determination of a renormalized (at least up to some finite 
number of loops) effective potential.

Finally, the analysis done here for a simple scalar field in a de 
Sitter background should be extended to a realistic case of the
fluctuations of the inflaton field coupled to the metric. This case
is of course more complicated because of the presence of a higher
number of fields and of the issue of gauge invariance, but the
conceptual basis of the procedure outlined here and the setup of the 
new diagrammatic formalism would not change.
The resulting ``renormalized'' cosmological perturbation theory 
would be of greatest interest, since for practical purposes it might
lead to different results. For example, the amount of non-Gaussianity 
hidden in loop contributions to the statistics of cosmological 
correlation functions might be larger than the standard prediction
obtained with tree-level terms only.
These aspects certainly deserve further investigation.

\section*{ACKNOWLEDGMENTS}

I am deeply thankful to Eiichiro Komatsu for many enlightening discussions 
and suggestions, as well as to Jonathan Rocher and Patrick Greene for
patiently answering to countless questions. It is also a pleasure to thank 
Steven Weinberg, J\'er\^ome Martin and Andrea Sartirana for reading the 
manuscript and giving helpful
comments. Part of this work has been done at the Institut d'Astrophysique
de Paris (IAP) and at the University of Milano-Bicocca, which are 
acknowledged for their hospitality.
This material is based upon work supported by the National Science
Foundation under Grant No. PHY-0455649.


\appendix

\section{More on anticommutators}
\label{app1}

As we have seen in Section \ref{graphrep}, anticommutators are 
introduced at each asymmetric vertex of an open graph, when the different 
branches have a different free field structure.
Symmetric vertices do not need any anticommutator: either the free fields 
are all evaluated at the same time (as in final vertices) or they can
be exchanged simply by renaming the integration variables.
However, when open graphs are combined into closed graphs, the symmetry
properties of each vertex also depend on what dashed lines are attached 
to it.

As an example, let us consider the third order contribution to the 
three-point function given by the contraction of the first term in
\eqref{grphi_3} with two free fields $\phi_0(x_2)$ and $\phi_0(x_3)$.
This has the expectation value
\begin{equation}
  \mean{\phi_0^2(z)\phi_0^2(w)\phi_0(x_2)\phi_0(x_3)}
\end{equation}
with no anticommutator in it (because we are free to exchange $z$ and 
$w$). However, the result is different depending on which external
free field is paired with which vertex. Actually, one gets
\begin{align}
  &4\mean{\phi_0(z)\phi_0(w)}\!\Big[\!\mean{\phi_0(z)\phi_0(x_2)}
  \mean{\phi_0(w)\phi_0(x_3)} \notag \\
  &+\mean{\phi_0(z)\phi_0(x_3)}\mean{\phi_0(w)\phi_0(x_2)}\!\Big];
\end{align}
after exchanging the variables, this becomes
\begin{equation}
  4\mean{\phi_0(z)\phi_0(x_2)}\mean{\phi_0(w)\phi_3}
  \big(\!\mean{\phi_z\phi_w}+\mean{\phi_w\phi_z}\!\big),
\label{intsymm}
\end{equation}
where the internal dashed line is now symmetrized. However, the
symmetrization does not follow from the presence of an anticommutator,
but from the sum of the two different pairings with $\phi_0(x_2)$ and 
$\phi_0(x_3)$.

The resulting closed graph representing this contribution is
\begin{equation}
  \begin{pic}{60}{40}{(-5,-20)}
    \SetWidth{0.5}
    \SetColor{Black}
    \Vertex(0,0){1.5}
    \ArrowLine(15,0)(0,0)
    \ArrowArc(25,0)(10,60,180)  \ArrowArcn(25,0)(10,300,180)
    \DashCArc(25,0)(10,300,60){3}
    \DashLine(30,8.66)(37.5,19.65){3}\Vertex(37.5,19.65){1.5}
    \DashLine(30,-8.66)(37.5,-19.65){3}\Vertex(37.5,-19.65){1.5}
    \Text(1,5)[]{\normalsize{\Black{$x_1$}}}
    \Text(12,-5)[]{\normalsize{\Black{$y$}}}
    \Text(28,14)[]{\normalsize{\Black{$z$}}}
    \Text(28,-14)[]{\normalsize{\Black{$w$}}}
    \Text(47,19)[]{\normalsize{\Black{$x_2$}}}    
    \Text(47,-21)[]{\normalsize{\Black{$x_3$}}}    
  \end{pic};
\end{equation}
even though in this case there is an intermediate vertex with no 
anticommutator associated, we can still follow the prescriptions 
outlined in Section \ref{Feyn} and interpret the internal dashed 
line as a symmetrized expectation value.
Therefore, we do not need a different rule to symmetrize internal 
dashed lines within a completely symmetric sub-tree, even though
the origin of the symmetrization is different.

We need anyway to pay attention to the symmetry factor: if we consider
the internal dashed line to be symmetric, there are 4 equivalent 
possibilities to pair the first external leg and 2 for the second. 
This would overestimate the correct number of graphs given in 
\eqref{intsymm} by a factor of 2. In general,
we must add a factor $\frac{1}{2}$ whenever the two solid branches to
which the internal dashed line(s) is (are) attached are completely 
symmetric.
This allows to adopt the same symmetrization rule for all internal dashed
lines, and at the same time to use the standard combinatorial techniques
when counting the number of equivalent graphs.

\end{document}